\documentclass[prl,aps,amssymb,amsmath,reprint,superscriptaddress]{revtex4-1}

\usepackage[T1]{fontenc} % Use Type1 fonts
\usepackage[utf8]{inputenc} % Set .tex encoding to utf8
\usepackage[UKenglish]{babel} % UK English hyphenation

\usepackage{booktabs} % Use booktabs and array for better tables
\usepackage{array} %q
\usepackage{graphicx}

\newcommand{\tr}{{\rm tr \thinspace}}
\newcommand{\bra}[1]{\left\langle{#1}\right\vert}
\newcommand{\ket}[1]{\left\vert{#1}\right\rangle}

\newcommand{\expect}[1]{\langle{#1}\rangle}

\newcommand{\beq}{\begin{equation}}
\newcommand{\eeq}{\end{equation}}
\newcommand{\bqa}{\begin{eqnarray}}
\newcommand{\eqa}{\end{eqnarray}}
\newcommand{\nn}{\nonumber}

\newcommand{\erf}[1]{Eq.~(\ref{#1})}
\newcommand{\dg}{^\dagger}

\newcommand{\im}[1]{\text{Im}\{{#1}\}}

\newcommand{\eg}{\emph{e.g.},~}
\newcommand{\ie}{\emph{i.e.},~}

\def\ko{\kappa_{{\rm s}, 1}}
\def\kt{\kappa_{{\rm s}, 2}}
\def\kio{\kappa_{{\rm w}, 1}}
\def\kit{\kappa_{{\rm w}, 2}}
\def\etal{\eta_{loss}}
\def\etam{\eta_{\rm meas}}

\def\be{\begin{equation}}
\def\ee{\end{equation}}
\def\bea{\begin{eqnarray}}
\def\eea{\end{eqnarray}}

\begin{document}

\author{N. Roch$^{*\dagger}$}
%\email[Corresponding author: ]{nicolas.roch@neel.cnrs.fr}
%\thanks{These authors contributed equally to this work.}
\affiliation{Quantum Nanoelectronics Laboratory, Department of Physics, University of California, Berkeley, California 94720, USA.}
\author{M. E. Schwartz$^{*}$}
%\thanks{These authors contributed equally to this work.}
\affiliation{Quantum Nanoelectronics Laboratory, Department of Physics, University of California, Berkeley, California 94720, USA.}
\author{F. Motzoi}
\affiliation{Department of Chemistry, University of California, Berkeley, California 94720, USA.}
\author{C. Macklin}
\affiliation{Quantum Nanoelectronics Laboratory, Department of Physics, University of California, Berkeley, California 94720, USA.}
\author{R. Vijay}
\affiliation{Department of Condensed Matter Physics and Materials Science, Tata Institute of Fundamental Research, Mumbai, 400005, India.}
\author{A. W. Eddins}
\affiliation{Quantum Nanoelectronics Laboratory, Department of Physics, University of California, Berkeley, California 94720, USA.}
\author{A. N. Korotkov}
\affiliation{Department of Electrical Engineering, University of California, Riverside, California 92521, USA.}
\author{K. B. Whaley}
\affiliation{Department of Chemistry, University of California, Berkeley, California 94720, USA.}
\author{M. Sarovar}
\affiliation{Scalable and Secure Systems Research (08961), Sandia National Laboratories, Livermore, CA 94550, USA.}
\author{I. Siddiqi}
\affiliation{Quantum Nanoelectronics Laboratory, Department of Physics, University of California, Berkeley, California 94720, USA.}

\title{Observation of measurement-induced entanglement and quantum trajectories of remote superconducting qubits}

%PACS codes are for: 03.67.Bg entanglement production; 42.50.Dv Quantum state engineering and measurements; 42.50.Lc Quantum ﬂuctuations, quantum noise, and quantum jumps; 85.25.Dq SQUIDs

\begin{abstract}
The creation of a quantum network requires the distribution of coherent information across macroscopic distances. We demonstrate the entanglement of two superconducting qubits, separated by more than a meter of coaxial cable, by designing a joint measurement that probabilistically projects onto an entangled state. By using a continuous measurement scheme, we are further able to observe single quantum trajectories of the joint two-qubit state, confirming the validity of the quantum Bayesian formalism for a cascaded system.  Our results allow us to resolve the dynamics of continuous projection onto the entangled manifold, in quantitative agreement with theory.

\end{abstract}

\date{\today}

\maketitle

 * These authors contributed equally to this work
\newline $\dagger$ Present address: CNRS and Universit\'e Grenoble Alpes, Institut N\'eel, 38042 Grenoble, France

Entanglement\textemdash the property that binds two independent objects into a single, highly correlated, nonseparable system\textemdash is a hallmark of quantum theory.  Entanglement schemes for superconducting qubits have traditionally relied on direct qubit-qubit coupling\cite{Steffen,Dewes}, cavity-mediated interactions\cite{DiCarlo}, photon-mediated interactions\cite{vanLoo} or  autonomous cooling\cite{Shankar}.  Measurement, in contrast, has traditionally been viewed as a means to restore classical behavior: a quantum system, once observed, is projected onto a single measurement basis state.  However, in certain cases it is possible to design\cite{Ruskov,Engel, Trauzettel,Hutchison, Lalumiere, Helmer} a measurement that projects onto an entangled state, thereby purifying, rather than destroying, quantum correlations. Such a measurement has recently been used to entangle two superconducting qubits coupled to the same microwave resonator\cite{Riste}.  

Measurement-induced entanglement is a particularly important resource in spatially-separated quantum systems, for which no local interactions and therefore no direct methods of creating entanglement exist.  Such remote entanglement has been demonstrated using optical photons in several atomic systems\cite{Chou, Hofmann,Moehring} and nitrogen vacancy centers\cite{Bernien}, but has remained elusive for superconducting qubits, which operate in the microwave regime.  In this Letter, we demonstrate measurement-induced entanglement between two superconducting qubits, each dispersively\cite{Wallraff} coupled to a separate cavity for readout and separated by 1.3 meters of ordinary coaxial cable, by engineering a continuous measurement for which one of the three outcomes is a Bell state\cite{Kerckhoff}.  Unlike previous experiments in spatially-separated quantum systems, in which the detection of individual spontaneous fluorescence events reveals whether or not entanglement has been generated, we employ time-continuous measurements\cite{Hatridge}.  This allows us to access the ensemble-averaged \textit{dynamics} of entanglement generation, which are well-described by a statistical model and by a full master-equation treatment.  Furthermore, our measurement efficiency is sufficiently high to resolve the individual quantum trajectories in the ensemble\cite{Murch}, thus enabling the observation of the stochastic evolution of a joint two-qubit state under measurement.  This functionality sheds new light on the fundamental interplay between entanglement, measurement, and decoherence in a quantum network.

Our experimental apparatus consists of two superconducting transmon qubits placed in spatially separated copper waveguide cavities (3D transmon architecture)\cite{Paik}.  Each cavity is wound with a superconducting bias coil to enable tuning of the qubit frequency.  A weakly coupled port is used for transmission measurements and single qubit control, and a strongly coupled port enables qubit state readout.  The strongly coupled ports of the two cavities are connected via two microwave circulators and 1.3 meters of coaxial cable to enable directional transfer of information from cavity 1 to cavity 2 (Figure 1a). The entire apparatus is contained within an absorptive shield and a Cryoperm magnetic shield to suppress spurious radiation and noise.  Qubit and cavity parameters are described in detail in the supplemental information\cite{supp}.  

\begin{figure}
\includegraphics[scale=0.32]{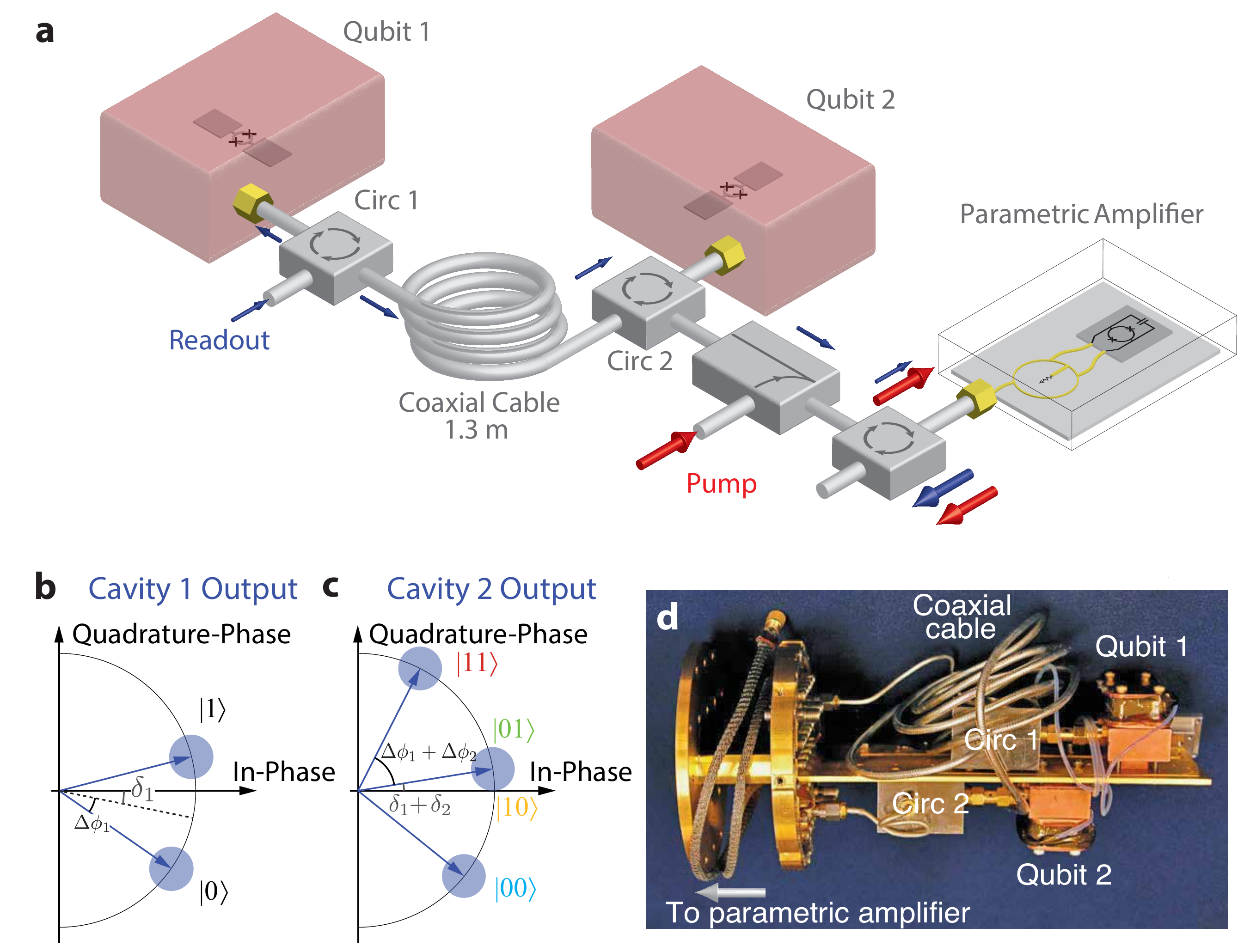}
\caption{Experimental setup. \textbf{a} Simplified representation of the experimental setup.  \textbf{b}, \textbf{c}  Schematic of the phase shift acquired by a coherent state sequentially measuring first qubit 1 (\textbf{b}) and then qubit 2 (\textbf{c}) in reflection. \textbf{d} Picture of the base-temperature setup.}
\end{figure}

A joint qubit state measurement can be performed by sequentially driving the cavities in reflection with a near-resonant microwave tone at frequency $\omega_m$ that can be described by a classical complex amplitude $\alpha_{in}$.  For a single qubit measured in reflection, the output state is given by $\alpha_{out} = r^{\pm}\alpha_{in}$, where the reflection coefficient $r^{\pm}$ is given by
\begin{equation}r^{\pm} = \frac{\kappa-2 i (\omega_r-\omega_m\pm\chi)}{\kappa + 2i(\omega_r - \omega_{m} \pm \chi) },\end{equation}
 and the signifier + (-) represents the single qubit state $\ket{0}$ $\left(\ket{1}\right)$\cite{supp}.  Here, $\omega_{r}$ is the bare cavity frequency; $\kappa$ is the cavity decay rate; and $\chi$ is the dispersive shift.  The measurement tone acquires a qubit state-dependent phase shift $\phi^{\pm}=\textrm{Arg} \left[\alpha_{out}^{\pm}\right]$. For the following analysis it is convenient to define the average and relative phase shifts, $\delta = \frac{1}{2} (\phi^++ \phi^-)$  and $\Delta \phi = \frac{1}{2} (\phi^+-\phi^-)$, respectively (See Figure 1b).

 For a sequential reflective measurement of two qubits, the output coherent state becomes $\alpha_{out} = \sqrt{\eta_{loss}}r^{\pm}_1 r^{\pm}_2 \alpha_{in}$, where $\eta_{loss}\approx 0.81$ represents the efficiency of power transfer between the two cavities.  In the general case, $\Delta\phi_1 \neq \Delta\phi_2$ and the phase shifts corresponding to the four basis states $\ket{00}$, $\ket{01}$, $\ket{10}$ and $\ket{11}$ are all distinct; the associated measurement decoheres any quantum superposition of states and projects the system into one of the four basis states. However, if we carefully engineer the cavities and the dispersive coupling\cite{supp}, there exists $\omega_m$ such that $\Delta\phi_1=\Delta\phi_2$. In this situation, the phase shifts associated with states $\ket{01}$ and $\ket{10}$ are identical and equal to $\delta_1+\delta_2$; the measurement therefore cannot decohere a quantum superposition of $\ket{01}$ and $\ket{10}$ (shown schematically in Figure 1c).   We use a superconducting parametric amplifier\cite{Hatridge2} to measure the acquired phase shift, realizing a high-fidelity homodyne measurement characterised by a quantum efficiency $\eta_{meas}=0.4\pm0.10$.  Figure 2a shows a sample time-domain trace of  the homodyne signal 
\begin{equation}V_m(t_m)=\frac{1}{t_m}\int_0^{t_m}V(t)dt,\end{equation}
where $V$ is the instantaneous voltage (inset).   We verify that our joint readout cannnot distinguish between $\ket{01}$ and $\ket{10}$ by sequentially preparing and then measuring the four basis states. Figure 2b represents histograms of $V_m$ for a measurement time $t_m=0.65$ $\mu $s. The states $\ket{00}$ and $\ket{11}$ are well-separated, while the histograms for $\ket{01}$ and $\ket{10}$ are fully overlapping, as desired.  This enables us to post-select measurement instances that correspond to occupation of the odd-parity manifold without destroying coherence within that manifold, and therefore to probabilistically generate entanglement.  

\begin{figure}
\begin{center}
\includegraphics[scale=0.5]{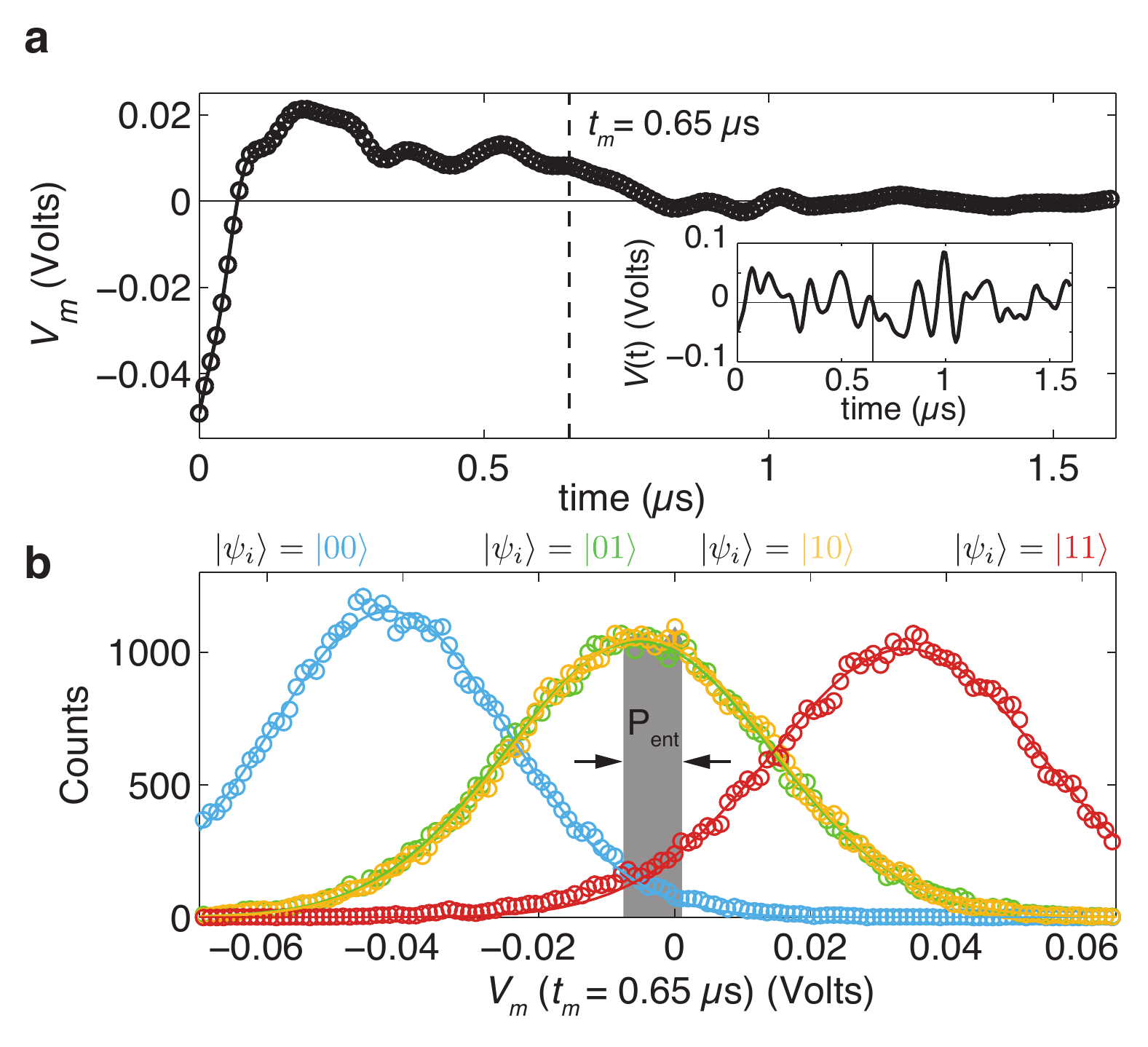}
\caption{Demonstration of indistinguishability between $\ket{01}$ and $\ket{10}$ computational states during measurement. \textbf{a} Example of the temporal evolution of the measurement signal $V_m$. The inset shows the associated instantaneous voltage $V(t)$. \textbf{b} Histogram of $V_m$ for each of the four computational states $\ket{00}$,  $\ket{01}$, $\ket{10}$ and $\ket{11}$. The range of data post-selected for tomographic reconstruction at $t_m=0.65 \mu s$ is represented as a shaded grey area.}
\end{center}
\end{figure}

\begin{figure}
\begin{center}
\includegraphics[scale=0.23 ]{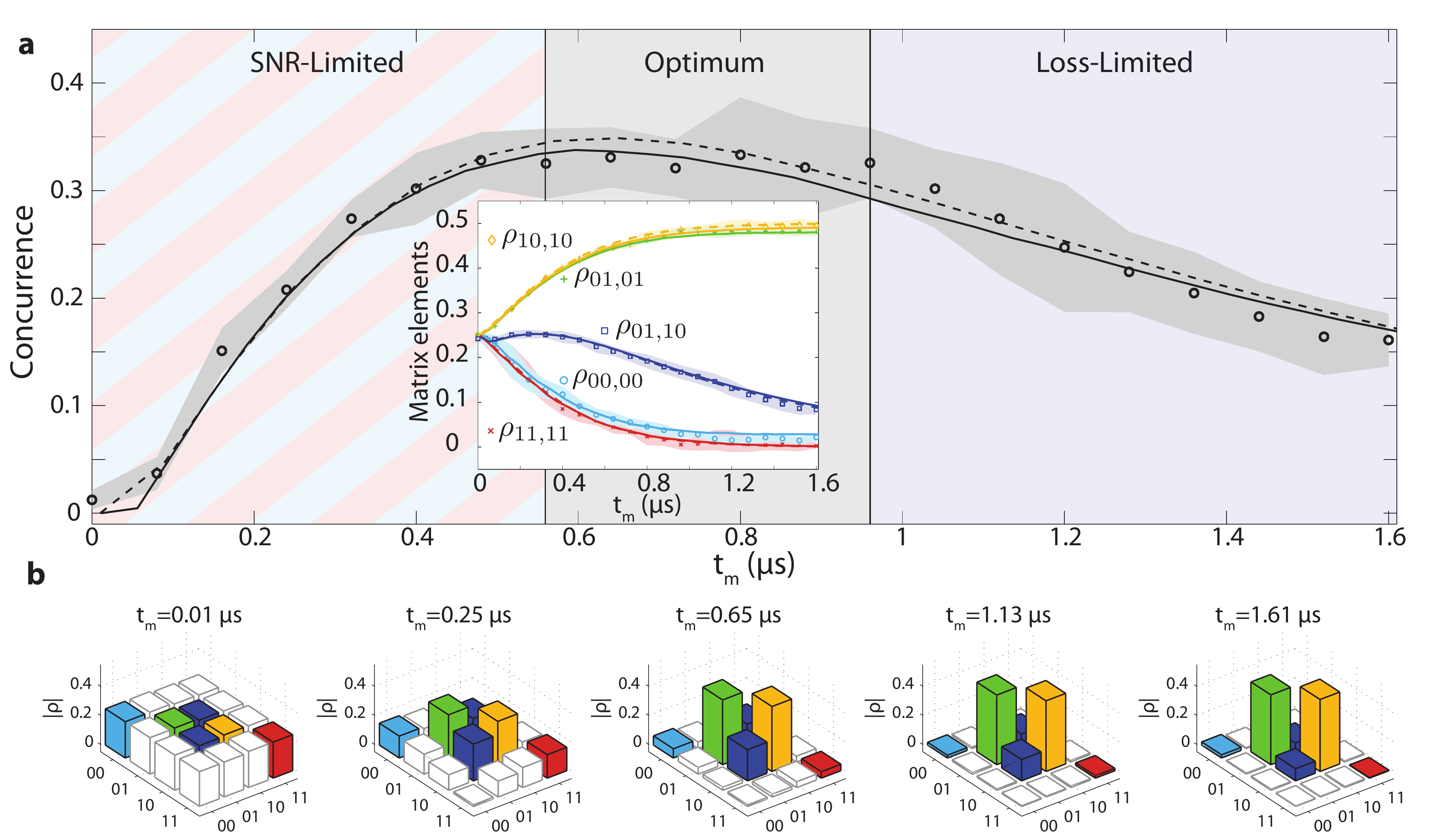}
\caption{Generation and verification of entanglement between two spatially-separated superconducting qubits. \textbf{a}  Concurrence of the entangled state as a function of $t_m$. The inset displays the evolution of the basis state populations ($\rho_{00,00}$, etc.) and odd-parity coherence $\left( \rho_{01,10} \right)$.  The shaded region represents the standard deviation centered about the average (circles). Dashed lines are theoretical simulations based on a Bayesian approach and solid lines are calculated using a rigorous master equation; in both cases no fitting parameter are used\cite{supp}. \textbf{b} Full density matrices of the post-selected entangled subspace for increasing $t_m$. }
\end{center}
\end{figure}

We control the rate of entanglement generation 
\begin{equation}\Gamma_{meas}=\frac{1}{2}\eta_{meas}\eta_{loss}|\alpha_{in}|^2\sin(2\Delta\phi)^2,\end{equation}
 by adjusting the measurement strength via the average intracavity photon number  $\overline{n}_1 = \frac{1}{2} \left( \overline{n}_1^+ + \overline{n}_1^- \right)$ where for each cavity $i$\cite{supp}
\begin{equation}\overline{n}_i^\pm = \frac{\kappa_{i}}{(\kappa_{i}/2)^2 + (\omega_{i} - \omega_{m} \pm \chi_{i})^2} |\alpha_{in}|^2.\end{equation}  
A photon number $\overline{n}_1=1.2$ results in $\Gamma_{meas}/2\pi \approx 210$ kHz, which sets the characteristic timescale of entanglement generation $\tau_{meas} \equiv 1/\Gamma_{meas}\approx 750$ ns.  Thus, the dynamics of the measurement process, which are significantly faster than qubit decay rates, can be readily resolved using conventional digital electronics. 

To generate and verify entanglement, we implement a sequence of three readout protocols and two qubit rotations.  We first perform a projective readout ($\overline{n}_1 =6.2$ and $1\ \mu$s readout length) to post-select the $\ket{00}$ ground state\cite{Johnson}.  We then perform $R^{\pi/2}_{y}$ rotations on both qubits to create the equal superposition state $\frac{1}{2} \left( \ket{00} + \ket{01} + \ket{10} + \ket{11} \right)$.  The second readout, which is done in the weak regime and with varying $t_m$, stochastically steers the system toward $\ket{00}$, $\ket{11}$, or the Bell state $\frac{1}{\sqrt{2}}(\ket{01}+\ket{10})$, as documented in the measurement output $V_m$. We then apply one of a set of 30 tomographic rotations immediately followed by a strong readout. We repeat this process 8,000 times for each tomographic rotation and for each $t_m$ to form a single well-averaged data set;  we generate an error margin by taking the average and standard deviation of 17 data sets. To produce the density matrix of the post-selected entangled state for each time $t_m$, we choose an entanglement probability $p_{ent}$ to constitute the entangled state based on $V_m(t_m)$  (shown in the grey shaded region in Figure 2b for $t_m=0.65$ $\mu s$),  and tomographically reconstruct the density matrix using a maximum-likelihood estimator\cite{supp}. For perfectly separated histograms, $50\%$ of the counts will lie in the odd-parity subspace, but we utilize $p_{ent}=10\%$ to compensate for imperfect measurement efficiency.  

The ability to perform time-continuous measurements enables us to directly observe the ensemble dynamics of the emergence of entanglement. Writing the two qubit density matrix as $\rho = \sum_{ijkl} \rho_{ij,kl} \ket{ij}\bra{kl}$, we can estimate concurrence \cite{Wootters} using the simplified formula \cite{Jakobczyk} ${\cal C} \approx \max(0,|\rho_{01,10}|-\sqrt{\rho_{00,00} \rho_{11,11}})$ to characterize the quality of the entanglement during this process. This simplified formula holds when the only non-negligible off-diagonal elements are $\rho_{01,10}$ and its conjugate, which is applicable to our setup since the high distinguishability between $\ket{00}, \ket{11}$ and the $\{\ket{01}, \ket{10}\}$ manifold results in rapid decay of all other off-diagonal elements. Concurrence ranges from zero (for a separable or mixed state) to one (for a  maximally entangled two qubit state), and is greater than zero for all non-separable two qubit states\cite{Wootters}.  Maximizing ${\cal C}$  requires  limiting decoherence within the odd-parity manifold, and minimizing stray counts of $\ket{00}$ and $\ket{11}$ by maximizing the signal-to noise ratio (SNR), defined by the ratio of the separation of the Gaussian histograms (in Figure 2b) to their width, or 
\begin{equation}\textrm{SNR}\sim2|\alpha_{in}|\sin\left(2\Delta\phi\right)\sqrt{\eta_{loss}\eta_{meas} t_m}.\end{equation} 
 Figure 3 shows the evolution of the concurrence as a function of $t_m$. The inset shows the evolution of the relevant density matrix elements (the diagonal elements, representing population probabilities, and the off-diagonal element $\rho_{01,10}$, representing the coherence of the odd-parity subspace). 

We note three qualitative regimes: SNR-dominated evolution; stabilization; and decay due to decoherence. Since SNR is proportional to $\sqrt{t_m}$, it dominates the evolution at short times $t_m < 0.75\tau_{meas}$.  Here, the dynamics are governed by changes to population probabilities; i.e., the increase of $\rho_{01,01}$ and $\rho_{10,10}$ and decrease of $\rho_{00,00}$ and $\rho_{11,11}$ in the post-selected ensemble.  The rapid decay of $\rho_{00,00}$ and $\rho_{11,11}$ compared to $\rho_{01,10}$, results in growth of concurrence in this regime. For intermediate times $\left(0.75\tau_{meas}<t_m<1.25\tau_{meas}\right)$, the SNR improvement rate decreases and decoherence begins to take a more noticeable effect.  Decoherence is caused by intrinsic dephasing of the qubits $\Gamma_{2,i}^* = 1/T_{2,i}^*$ and by $\eta_{loss}$, which contributes an additional measurement-induced dephasing of the first qubit at a rate 
\begin{equation}\Gamma_{loss}\simeq 2\left(1-\eta_{loss}\right)|\alpha_{in}|^2\sin(\Delta\phi)^2.\end{equation}
At intermediate times, the SNR improvement rate and $\Gamma_{loss}$ are roughly equal, and hence the concurrence reaches a maximum value of 0.35. This value is comparable to what was obtained recently using optical communications\cite{Moehring,Bernien}, however, thanks to our time-continuous measurement scheme, the rate at which a qubit-qubit entangled state is created is orders of magnitude higher ($\Gamma_{creation}/2\pi=1 \textrm{ kHz}$).  For longer times $\left(t_m > 1.25 \tau_{meas}\right)$, the density matrix evolution is dominated by decoherence, which eventually drives the system into an incoherent mixture of $\ket{01}$ and $\ket{10}$. 

These ensemble dynamics are well-described both by a simple statistical model (dashed lines), and by a rigorous master-equation treatment (solid lines)\cite{supp}. The models, which account for the chief technical limitations of our scheme (i.e. the inefficiencies  $\eta_{loss}$, the losses between the cavities and $\eta_{meas}$, the finite detection efficiency), indicate that reasonable technical improvements could lead to concurrence of $70\%$, which is comparable to recent single cavity experiments\cite{Riste}.

\begin{figure}
\begin{center}
\includegraphics[scale=0.3]{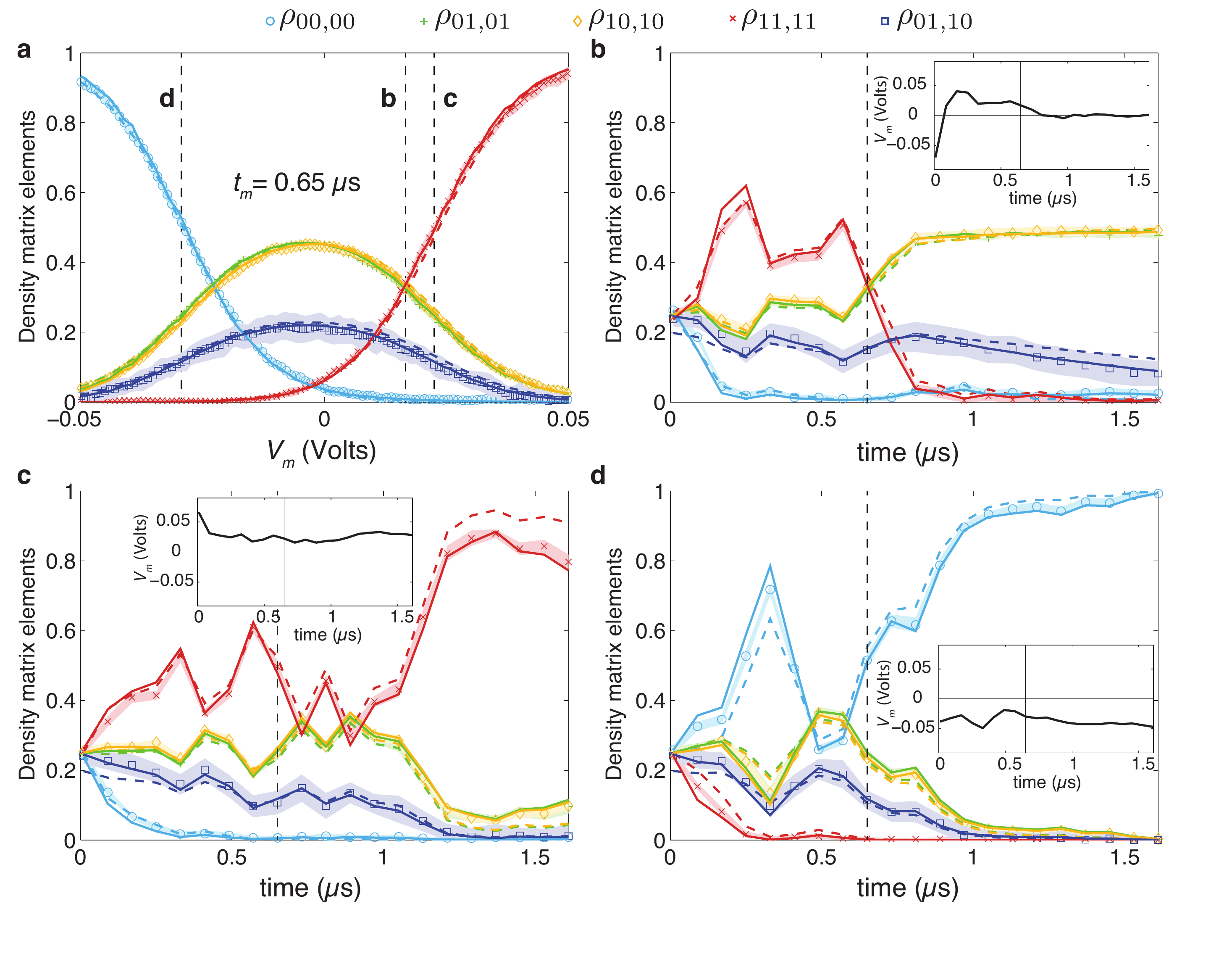}
\caption{Resolving single quantum trajectories for cascaded quantum systems. \textbf{a} Absolute value of the density matrix elements conditioned on the measured voltage $V_m$ for $t_m=0.65\ \mu s$ and $\overline{n}_1=1.2$, presenting an instantaneous mapping $V_m \mapsto \rho \left(  V_m \right)$.  The shaded region represents the standard deviation about the average (circles); dashed lines (resp. solid lines) are theoretical simulations based on a Bayesian approach (resp. on a full master equation) without fitting parameter\cite{supp}. \textbf{b},\textbf{c},\textbf{d} Examples of reconstructed quantum trajectories for diagonal and principal off-diagonal density matrix elements. The dots represent tomographic reconstruction based on the mapping $V_m \mapsto \rho \left(  V_m \right)$ for every $t_m$. The dashed lines are Bayesian estimations based on the measured $V_m(t)$ (insets).The solid lines for the full master equation were obtained by running 100000 instances of the stochastic differential equation with 1ns resolution and averaging the obtained populations conditioned on $V_m$ at $t_m$\cite{supp}.
}
\end{center}
\end{figure}

Our high-efficiency continuous measurement allows us to go one step further in decomposing the dynamics of measurement-induced entanglement: we can directly observe the individual quantum trajectories\cite{Murch, Campagne} of our two qubit system, using a Bayesian update process.  In this formalism, $V_m(t)$ contains partial quantum state information that allows us to update our estimate of the instantaneous quantum state of the two-qubit system.  To validate the Bayesian update for a single trajectory, we generate a mapping $V_m \mapsto \rho \left(  V_m \right)$: at each time $t_m$, we collect all trajectories with similar weak measurement outcomes, $V(t_m)$, and perform a conditional tomographic state reconstruction of those trajectories (see Figure 4a). We then use this mapping to convert the measured voltage $V_m(t)$ for a \textit{single} experimental realization into $\rho(t)$ and thus reconstruct the quantum trajectory of the system\cite{Murch}. Figure 4b illustrates three typical trajectories, in which the system is projected onto a Bell state or onto the non-entangled states $\ket{00}$ or $\ket{11}$. We see excellent agreement between the tomographic reconstructions of the trajectories and theoretical predictions based on Bayesian updates and a master equation treatment. The observation of these quantum trajectories shows the novelty and strength of our continuous measurement scheme. Our experiment thus demonstrates the validity of quantum trajectory theories for cascaded quantum systems \cite{Carmichael, Gardiner}, which describe the conditioned evolution of distributed quantum systems.

Our experiments demonstrate that quantum entanglement can be established between distant systems that interact only through a coherent signal propagating along low loss electrical wires, a functionality that will be integral to the realization of complex, distributed quantum networks.  We take advantage of the versatility of continuous measurement to monitor the dynamics of entanglement generation, and demonstrate quantitative agreement to a theoretical model that captures the experimental details of the physical circuit\cite{supp}. Moreover, our characterization of the state of the joint system under continuous measurement suggests the feasibility of future \textit{continuous} feedback stabilization of entanglement \cite{Sarovar,Hofer}. Further technical improvements in quantum efficiency, coherence times, and transmission characteristics hold the promise of on-demand, stabilized remote entanglement\textemdash a powerful resource for quantum information processing.

\begin{acknowledgments}
We thank E. M. Levenson-Falk, K. W. Murch, D. H. Slichter, D. M. Toyli and S. J. Weber for discussions. This research was supported in part by the US Army Research Office, the Intelligence Advanced Research Projects Activity (IARPA).  MES acknowledges support from the Fannie and John Hertz Foundation.  RV acknowledges support from the Government of India. ANK acknowledges support from ARO under MURI W911NF-11-1-0268. Sandia National Laboratories is a multi-program laboratory managed and operated  by Sandia Corporation, a wholly owned subsidiary of Lockheed Martin Corporation, for the United States Department of Energy's National Nuclear Security Administration under contract DE-AC04-94AL85000.  
\end{acknowledgments}

%\bibliography{miebiblio}

\newpage
\setcounter{figure}{0}
\setcounter{table}{0}
\setcounter{equation}{0}
\onecolumngrid

\global\long\def\theequation{S\arabic{equation}}

\global\long\def\thefigure{S\arabic{figure}}

\vspace{1.0cm}
\begin{center}
{\bf \large Supplementary Information for "Observation of measurement-induced entanglement and quantum trajectories of remote superconducting qubits"}
\vspace{0.5cm}\\

\end{center}

\renewcommand{\figurename}{Fig S}

%\author{Nicolas Roch$^{*\dagger}$}
%\author{Mollie E. Schwartz$^{*}$} 
%\affiliation{Quantum Nanoelectronics Laboratory, Department of Physics, University of California, Berkeley, California 94720, USA.} 
%\author{Felix Motzoi}
%\affiliation{Department of Chemistry, University of California, Berkeley, California 94720, USA.}
%\author{Christopher Macklin}
%\affiliation{Quantum Nanoelectronics Laboratory, Department of Physics, University of California, Berkeley, California 94720, USA.}
%\author{Rajamani Vijay}
%\affiliation{Department of Condensed Matter Physics and Materials Science, Tata Institute of Fundamental Research, Mumbai,400005, India.}
%\author{Andrew W. Eddins}
%\affiliation{Quantum Nanoelectronics Laboratory, Department of Physics, University of California, Berkeley, California 94720, USA.}
%\author{Alexander N. Korotkov}
%\affiliation{Department of Electrical Engineering, University of California, Riverside, California 92521, USA.}
%\author{Birgitta Whaley}
%\affiliation{Department of Chemistry, University of California, Berkeley, California 94720, USA.}
%\author{Mohan Sarovar}
%\affiliation{Scalable and Secure Systems Research (08961), Sandia National Laboratories, Livermore, CA 94550, USA}  
%\author{Irfan Siddiqi}
%\affiliation{Quantum Nanoelectronics Laboratory, Department of Physics, University of California, Berkeley, California 94720, USA.}
%
%
%
% * These authors contributed equally to this work
%\newline $\dagger$ Present address: CNRS and Universit\'e Grenoble Alpes, Institut N\'eel, 38042 Grenoble, France

\section{Simplified theory}

In this section we describe a simplified phenomenological theory for calculation of the concurrence in our experiment. This theory is not intended to give a rigorously accurate result, but can be used for quick estimates and for gaining physical intuition. For clarity, we use $A(t)$ and $B(t)$ to represent the intracavity fields in cavity 1 and cavity 2, respectively, and $A_{out}$, $B_{out}$ to represent the propagating fields travelling \textit{from} the respective cavities.
\\
\\
For simplicity we neglect the off-diagonal elements $\rho_{00,01}$, $\rho_{00,10}$, $\rho_{11,01}$, and $\rho_{11,10}$ of the two-qubit density matrix (which should be small in the interesting regime -- see below), so that we have the so-called X-state and therefore can use the simplified formula \cite{Jacobczyk} for the concurrence,
       \be
    {\cal C} =  2 \max (0, |\rho_{01,10}| -
    \sqrt{\rho_{00,00} \rho_{11,11}} ),
    \label{concurrence-simple}\ee
which depends only on two diagonal elements and one off-diagonal element of the density matrix. (This concurrence does not depend on the element $\rho_{11,00}$, so we do not have to neglect it; however, in experiment it is even smaller than the neglected elements.)
\\
\\
To find $\rho_{00,00}$, $\rho_{11,11}$, and $\rho_{01,10}$ after the measurement, we first consider the case without energy relaxation and intrinsic (not measurement-induced) dephasing of the qubits; then the dynamics of the two-qubit state are only due to measurement. For each of four "classical'' initial
states of the qubits ($|00\rangle$, $|01\rangle$, $|10\rangle$, $|11\rangle$) it is easy to calculate the evolution of the classical field amplitudes $A(t)$ and $B(t)$ in the first and second resonators,
    \begin{eqnarray}
&&    \dot{A}=-\frac{\kappa_1}{2} A - i (\omega_{r,1} \pm \chi_1-\omega_m)A +
    \sqrt{\kappa_{s,1}} \, A_{d}(t),
    \label{A-dot}\\
&&    \dot{B}=-\frac{\kappa_2}{2} B - i (\omega_{r,2} \pm \chi_2 -\omega_m)B + \sqrt{\kappa_{s,2}}\sqrt{\eta_{loss}}\,
A_{out}(t),
    \label{B-dot}\end{eqnarray}
As in the main text, $+$ $(-)$ refers to the qubit state $|0\rangle$ $\left( |1\rangle\right)$.  Here the rotating frame ($e^{-i\omega_m t}$) is based on the measurement drive, the time for $B(t)$ is shifted by the "flying time'' between resonators, $A_{d}(t)$ is the external microwave drive amplitude ($\alpha_{in}$ in the main paper, $\kappa_1=\kappa_{s,1}+\kappa_{w,1}+\kappa_{decay,1}$ is the total bandwidth of the first resonator (including the bandwidth due to strongly and weakly coupled ports \textemdash see Fig.\ref{figure1}), and similarly $\kappa_2=\kappa_{s,2}+\kappa_{w,2}+\kappa_{decay,2}$ for the second resonator. The energy decay for the microwave propagation {\it between} the resonators is described by the efficiency $\eta_{loss}$, which describes the losses in the circulator and microwave cables. Notice that in Eqs.\ (\ref{A-dot}) and (\ref{B-dot}) the resonator field amplitudes $A$ and $B$ are normalized such that $|A|^2$ and $|B|^2$ are equal to the average number of photons in the corresponding coherent states, while for the propagating field $A_d$ the squared amplitude $|A_d|^2$ is equal to the average number of photons per unit time.
\\
\\
Similar normalization is used for the propagating field
\begin{equation}
A_{out}(t)= -A_{d}(t)
+\sqrt{\kappa_{s,1}}\,A(t),
\label{A-out}\\
\end{equation}
and the field
\begin{equation}
B_{out}(t)= -\sqrt{\eta_{loss}}\,A_{out}(t)
+\sqrt{\kappa_{s,2}} \, B(t),
    \label{B-out}\end{equation}
which goes from the second resonator through the circulator to the amplifier. In the steady-state limit ($\dot A=0$) and for $\kappa_{s}\gg\kappa_{w} + \kappa_{decay}$, we recover the experessions given in the main paper for the reflection coefficient:
\begin{equation}
r^{\pm}=\frac{\kappa_s-2i(\omega_r-\omega_m\pm\chi)}{\kappa_s+2i(\omega_r-\omega_m\pm\chi)},
\label{coeffref}
\end{equation}
and for the photon number population inside cavity 1:
\begin{equation}
\bar n^{\pm}_1=\frac{\kappa_{s,1}}{ \left(\kappa_{s,1}/2\right)^2 + \left( \omega_1 - \omega_m \pm \chi_1\right)^2} \left| A_{d} \right|^2.
\end{equation}
\\
\\
 To produce the entangled state in our experiment, the steady-state fields $B_{out}^{(01)}$ and  $B_{out}^{(10)}$ for the states $|01\rangle$ and $|10\rangle$ should be  indistinguishable, $B_{out}^{(01)}= B_{out}^{(10)}$, while they should be sufficiently well distinguishable from the fields
$B_{out}^{(00)}$ and  $B_{out}^{(11)}$. For amplification and homodyne measurement of the field quadrature $e^{i\phi}$, the average time-integrated measurement result for the state $|ij\rangle$ is
    \be
  S_{ij}=\frac{1}{t}\int {\rm Re} [ B_{out}^{(ij)}(t') \, e^{-i\phi}] \,
   f_w(t') \, dt',
    \label{S_ij}\ee
where $f_w(t)$ is the weight function (in the experiment we used constant-weight integration with adjustable start/end time moments).  The amplifier noise is also accumulated during this time-integration, so that for the two-qubit state $|ij\rangle$ the random measurement result is characterized by the Gaussian distribution with the mean value of $S_{ij}$ and the standard deviation
    \be
    \sigma = \frac{1}{2\sqrt{\eta_{meas}}} \sqrt{\frac{1}{t}\int f_w^2(t)\, dt},
    \label{sigma}\ee
where $\eta_{meas}$ is the quantum efficiency of the measurement setup, which includes quantum efficiency of the phase-sensitive amplifier and losses in the circulators and cables. Notice that the noise $\sigma$ does not depend on the two-qubit state. In our experiment $\eta_{meas}= 0.4$, $\eta_{loss}= 0.75$, and the measured phase $\phi$ is chosen to be perpendicular to the output states for $|01\rangle$ and $|10\rangle$, $\phi= {\rm arg}(B_{out}^{(10)}) = {\rm arg} (B_{out}^{(01)})$.
\\
\\
In the experiment we select only realizations for which the integrated signal falls within a certain range, centered near $(S_{01} + S_{10} )/2$. The total probability of selection in our model (assuming no energy relaxation of qubits) is then
    \be
    p_{ent} =\sum_{i,j} \rho_{ij,ij}^{in} \, p_{sel}(i,j),
    \ee
where $\rho^{in}$ is the two-qubit density matrix before the measurement and $p_{sel} (i,j)$ is the selection probability for the initial state $|ij\rangle$ (it is equal to the integral, within the selection range, of the Gaussian with mean value $S_{ij}$ and standard deviation $\sigma$). In the experiment the selection range is typically chosen to keep 10\% of realizations, $p_{ent}=0.1$.
\\
\\
Since the two-qubit state evolution is only due to measurement, the diagonal matrix elements of the final density matrix $\rho^{fin}$ should obey \cite{Korotkov-Bayes} the classical Bayes rule
    \be
    \rho^{fin}_{ij,ij} = \frac{\rho_{ij,ij}^{in}\,
    p_{sel}(i,j)}{p_{ent}}.
    \label{rho_fin} \ee
For the main off-diagonal matrix element $\rho^{fin}_{01,10}$ needed to calculate concurrence, the quantum Bayesian approach \cite{Korotkov-Bayes} cannot be applied rigorously; however, we can modify it phenomenologically by using the following approximation:
    \begin{eqnarray}
&& \hspace{-0.7cm}    |\rho_{01,10}^{fin}|=  |\rho_{01,10}^{in}| \,
    \frac{\sqrt{\rho_{01,01}^{fin}\rho_{10,10}^{fin}}}
    {\sqrt{\rho_{01,01}^{in}\rho_{10,10}^{in}}}
    \nonumber \\
&&  \times    \exp \left[
    -\frac{1}{2}\int |B_{out}^{(01)}(t)-B_{out}^{(10)}(t)|^2
    dt
    \right]
    \nonumber \\
  && \times \exp \left[
    -\frac{1}{2}\int\left(\left(1-\eta_{loss}\right) \kappa_{s,1}+\kappa_{w,1} + \kappa_{decay, 1}\right)  |A^{(01)}(t)-A^{(10)}(t)|^2
    dt
    \right]
  \nonumber\\
&& \times \exp \left[
    -\frac{1}{2}\int \left( \kappa_{w,2} + \kappa_{decay,2} \right) \, |B^{(01)}-B^{(10)}|^2
    dt
    \right] , \quad
    \label{rho-01,10-num}\end{eqnarray}
where the last three factors describe the dephasing due to potential distinguishability of states $|01\rangle$ and $|10\rangle$ in the field $B_{out}$ and "lost'' fractions of the fields $A$ and $B$ from the first and second resonators. The form of these dephasing factors directly follows from the overlap between two coherent states $|A_1\rangle$ and $|A_2\rangle$ in a resonator \cite{Walls-Milburn}: $ |\langle A_1|A_2\rangle | =\exp(-|A_1-A_2|^2/2)$.
\\
\\
Only the absolute value of $\rho_{01,10}^{fin}$ is needed to calculate the concurrence (\ref{concurrence-simple}). For completeness, the phase change of $\rho_{01,10}$ due to measurement can be approximately calculated using the master equation result \cite{Gambetta-08}
    \begin{equation}
\arg(\rho_{01,10}^{fin})-\arg(\rho_{01,10}^{in})=
 2\chi_1 \int {\rm Re}[A^{(01)}(t)A^{(10)}(t)^*] \, dt
 - 2\chi_2 \int {\rm Re} [B^{(01)}(t) B^{(10)}(t)^*] \, dt.
    \label{phase-01,10-2} \end{equation}
(Here we used a frame that takes care of unequal bare frequencies of the qubits.)
\\
\\
Now let us discuss the density matrix element $\rho_{00,01}^{fin}$, which was neglected in the calculation of concurrence (\ref{concurrence-simple}). Very crudely, it can be estimated as $|\rho_{00,01}^{fin}|\alt \sqrt{\rho_{00,00}^{fin}\rho_{01,01}^{fin}} \exp[-\frac{1}{2}(1-\eta_{meas})\int |B_{out}^{(00)}-B_{out}^{(01)}|^2\, dt ]$, where the exponential term is due to the "unmeasured'' part of $B_{out}$. In the interesting regime (when a significant entanglement is achieved) we have $\rho_{00,00}^{fin}\ll 1$ and the exponential term is also small because distinguishability of the states $|00\rangle$ and $|01\rangle$ is governed by a similar factor. This is why $\rho_{00,01}^{fin}$ is strongly suppressed, and we believe it can be neglected in approximate calculation of concurrence. Similar arguments can be used to show strong suppression of the density matrix elements $\rho_{00,10}^{fin}$, $\rho_{11,01}^{fin}$, and $\rho_{11,10}^{fin}$ in the regime interesting for producing significant entanglement.
\\
\\
So far we have assumed absence of intrinsic decoherence of the qubits. Pure dephasing of the qubits with the corresponding dephasing time $T_{\varphi,1}$ and $T_{\varphi,2}$ can be easily included into the calculation of concurrence by multiplying the main off-diagonal element $\rho_{01,10}^{fin}$ by the factor $\exp(-t_m/T_{\varphi,1}-t_m/T_{\varphi,2})$, where $t_m$ is the total duration of the measurement procedure. Including the energy relaxation is not so easy, but since its contribution is quite small in the experiment, this can be done in a very crude way. For example, instead of the energy relaxation occuring during the measurement, we can phenomenologically introduce the energy relaxation for time $t_{before}$ before the measurement and then for time $t_{after}$ after the measurement. A better way can be realized by assuming energy decay at a specific random time, and then adding two corresponding parts of the signal integration (\ref{S_ij}); however, this complication does not seem necessary for our simplified theory.

%%%%%%%%%%%%%%%%%%%%%%Theory Section%%%%%%%%%%%%%%%%%%%%%%%%%%%

\section{Theoretical model based on quantum trajectory theory}

A sequential probe of two cavities as in Fig. 1a of the main text is often referred to as a cascaded systems setup, and Carmichael \cite{Car-1993} has developed the quantum trajectory equations describing such one-way sequential probes of cascaded systems. Following this work we can write a stochastic master equation (SME) model for the experimental setup that includes qubit and cavity degrees of freedom:
\bqa
\frac{\text{d}\rho}{\text{d}t} &=& -i[H, \rho] + \mathcal{D}[\sqrt{\ko(1-\etal)} a]\rho + \mathcal{D}[-\sqrt{\ko\etal} a  + \sqrt{\kt}b ]\rho  + \kio \mathcal{D}[a]\rho  + \kit \mathcal{D}[b]\rho \nn \\
&& + \sqrt{\etam}\xi(t) \mathcal{H}[e^{i\phi}(-\sqrt{\ko\etal} a + \sqrt{\kt}b)]\rho \nn \\
&& +  \sum_{i=1}^2 \Gamma_\varphi^i \mathcal{D}[\sigma_z^i]\rho +  \sum_{i=1}^2 \Gamma_{r }^i \mathcal{D}[\sigma_-^i]\rho
\label{eq:sme}
\eqa
with variables as defined in the previous section. This equation is in Ito form and therefore $\xi(t)dt = dW(t)$. $dW(t)$ is a Wiener increment satisfying $E\{dW(t)\}=0$ and $E\{dW(t)dW(s)\}=\delta(t-s)$ ($E\{\cdot\}$ denotes expectation value). $a (b)$ is an annihilation operator for the intracavity field in cavity 1 (2). $\sigma_\alpha^{1(2)}$ is the $\alpha$ Pauli operator for qubit 1 (2). The superoperators above are defined as: $\mathcal{D}[A]B \equiv ABA\dg - \frac{1}{2} A\dg A B - \frac{1}{2} B A\dg A$ and $\mathcal{H}[A]B \equiv AB + BA\dg -\tr(AB+BA\dg)B$. The last line in this equation is the dephasing and relaxation of the qubits, and we assume these are described by Markovian processes -- \eg $\Gamma_{r}^i = 1/T_1^i$. This equation describes the conditioned state of the system under a measurement voltage trace
\beq
V(t) = \sqrt{\etam}\langle -\sqrt{\ko\etal} a  + \sqrt{\kt}b \rangle + \xi(t)
\eeq
The observable that is being monitored is $-\sqrt{\ko\etal} a  + \sqrt{\kt}b$ in terms of the intra-cavity field operators. Note that an equivalent way to write the above SME is to replace $\xi(t)$ with the quantity $V(t) - \sqrt{\etam}\langle -\sqrt{\ko\etal} a  + \sqrt{\kt}b \rangle$, which is the difference between what is measured and the best estimate of the observable. 
\\
\\
The transmission time between the two cavities is taken to be negligible and therefore a direct coupling effective Hamiltonian between intracavity fields can be derived using the methods in \cite{Car-1993}. This effective Hamiltonian is
\bqa
H &=& -\frac{\omega_{q,1}}{2}\sigma_z^1 -\frac{\omega_{q,2}}{2} \sigma_z^2 + \omega_{r,1} a\dg a + \omega_{r,2} b\dg b + \chi_1 a\dg a \sigma_z^1 + \chi_2 b\dg b \sigma_z^2  \nn \\
&&- i \frac{\sqrt{\ko \kt \etal}}{2}(a\dg b - b\dg a ) + i A_{d}(t) \sqrt{\kio} a\dg - i A_{d}^*(t) \sqrt{\kio}a 
\eqa
where $\omega_{q,i}$ is the qubit transition frequency.  This Hamiltonian is in the rotating frame with respect to the measurement tone frequency -- i.e rotating frame with respect to $H_0 = \omega_{m} a\dg a + \omega_{m} b\dg b$. The coupling between cavities is mediated by a propagating field and therefore is irreversible. The combination of the Hamiltonian and dissipative components of \erf{eq:sme} result in a unidirectional coupling, as will be seen below. 
\\
\\
In the following we do not consider driving of the qubit states and assume that the qubit states are $\sigma_x$ eigenstates at $t=0$. Simultaneous modeling of the projective dynamics of the measurement and qubit driving is challenging and must be done by a careful adiabatic elimination \cite{Gam.Bla.etal-2008,Motzoi:2013}. We find that modeling the additional dynamics introduced by the interplay of these two aspects is not necessary to get a good match to experiment and therefore consider a perfectly prepared initial state.
\\
\\
The Heisenberg equations of motion for expected values of the intra-cavity fields under unconditioned evolution (the unconditioned/average evolution is the same as \erf{eq:sme} but without the stochastic last term) are:
\bqa
\dot{\expect{a}} &=& -i \omega_{r,1} \expect{a} - i\chi_1 \expect{\sigma_z^a a} - \frac{\ko + \kio}{2} \expect{a} + A_{d}(t) \sqrt{\kio} \\
\dot{\expect{b}} &=& -i \omega_{r,2} \expect{b} - i\chi_2 \expect{\sigma_z^b b}  - \frac{\kt + \kit}{2} \expect{b} + \sqrt{\ko \kt \etal} \expect{a}
\eqa
These evolution equations make explicit the fact that the second cavity is driven by the first but not vice-versa. From these equations we can write evolution equations for coherent states of the intra-cavity fields conditioned on the qubits being in given states:
\bqa
\dot{A}^{(0)} &=& -i\omega_{r,1} A^{(0)} - i \chi_1 A^{(0)} - \frac{\ko+\kio}{2} A^{(0)} + A_{d}(t) \sqrt{\ko} \nn \\
\dot{A}^{(1)} &=& -i\omega_{r,1} A^{(1)} + i \chi_1 A^{(1)} - \frac{\ko+\kio}{2} A^{(1)} + A_{d}(t) \sqrt{\ko} \nn \\
\dot{B}^{(11)} &=& -i\omega_{r,2} B^{(11)} + i\chi_2 B^{(11)} - \frac{\kt + \kit}{2} B^{(11)}  + \sqrt{\ko \kt \etal} ~A^{(1)} \nn \\
\dot{B}^{(10)} &=& -i\omega_{r,2} B^{(10)} - i\chi_2 B^{(10)} - \frac{\kt + \kit}{2} B^{(10)}  + \sqrt{\ko \kt \etal} ~A^{(1)} \nn \\
\dot{B}^{(01)} &=& -i\omega_{r,2} B^{(01)} + i\chi_2 B^{(01)} - \frac{\kt + \kit}{2} B^{(01)}  + \sqrt{\ko \kt \etal} ~A^{(0)} \nn \\
\dot{B}^{(00)} &=& -i\omega_{r,2} B^{(00)} - i\chi_2 B^{(00)} - \frac{\kt + \kit}{2} B^{(00)}  + \sqrt{\ko \kt \etal} ~A^{(0)} 
\label{eq:cav_eqns}
\eqa
where $A = \expect{a}, B = \expect{b}$ and the superscripts indicate the conditioning on qubit states. The state of the second cavity is conditioned on the states of both qubits but the state of the first cavity is only conditioned on the first qubit state since there is no information flowing from the second to the the first cavity. In other words, $A^{(11)}=A^{(10)}=A^{(1)}$ and $A^{(01)}=A^{(00)}=A^{(0)}$. The probe field "bounces" off both cavities and the resulting output field that is measured, in terms of these intra-cavity fields, is:
\beq
B_{out}(t) = -\sqrt{\kappa_{s,1} \eta_{loss}}A(t) + \sqrt{\kappa_{s,2}}B(t)
\eeq
\\
\\
In Ref. \cite{Motzoi:2013} we generalize the techniques developed for a single qubit in a cavity in Ref. \cite{Gam.Bla.etal-2008} to the case relevant here of two cavities with embedded qubits. This generalization allows us to eliminate the cavity degrees of freedom and obtain an equation of motion just for the qubits that aids in assessing the performance of the remote entanglement scheme. However, for the purposes of modeling the present experiment we only detail part of the calculation.
\\
\\
The dynamical equation in \erf{eq:sme} is sufficient to model the experiment, however it is difficult to simulate since it involves both qubit and cavity degrees of freedom. Instead, we will derive an effective SME for the qubit degrees of freedom only. The first step is to perform a polaron transformation into a frame where the average state of both cavities is the vacuum. The correct transform in this two cavity case is $\rho^P(t) = U(t)\dg \rho(t) U(t)$, with
\bqa
U(t) &= \Pi_{11}D_1\left[A^{(1)}(t)\right]D_2\left[B^{(11)}(t)\right] +  \Pi_{10}D_1\left[A^{(1)}(t)\right]D_2\left[B^{(10)}(t)\right] \nn \\
&+  \Pi_{01}D_1\left[A^{(0)}(t)\right]D_2\left[B^{(01)}(t)\right] +  \Pi_{00}D_1\left[A^{(0)}(t)\right]D_2\left[B^{(00)}(t)\right]
\eqa
where $\Pi_{ij} = \ket{i}_1\bra{i}\otimes \ket{j}_2\bra{j}$ are projectors onto qubit states and $D_{1(2)}[X]$ is a displacement operator for cavity field $1 (2)$ by coherent state $X$. In this frame, the equation of motion (for an unnormalized density matrix in the polaron frame) becomes
\bqa
\frac{\text{d}\rho^P}{\text{d}t} &=& -i[H^P + \frac{i\sqrt{\kio}}{2}A_{d}^*(t)\Pi_a -  \frac{i\sqrt{\kio}}{2}A_{d}(t)\Pi_a\dg  ] \nn \\
&& + \mathcal{D}[\sqrt{\ko(1-\etal)} a]\rho^P + \mathcal{D}[-\sqrt{\ko\etal} a  + \sqrt{\kt}b ]\rho^P  + \kio \mathcal{D}[a]\rho^P  + \kit \mathcal{D}[b]\rho^P \nn \\
&& + \mathcal{D}[\sqrt{\ko(1-\etal)} \Pi_a]\rho^P + \mathcal{D}[-\sqrt{\ko\etal} \Pi_a  + \sqrt{\kt}\Pi_b ]\rho^P  + \kio \mathcal{D}[\Pi_a]\rho^P  + \kit \mathcal{D}[\Pi_b]\rho^P \nn \\
&& + \Gamma_1\big( a[\rho^P, \Pi_a\dg] + [\Pi_a, \rho^P]a\dg \big) + \Gamma_2 \big( b[\rho^P, \Pi_b\dg] + [\Pi_b, \rho^P] b\dg \big) \nn \\
&&+ \sqrt{\ko\kt\etal} \big( a[\rho^P, \Pi_b\dg] + [\Pi_b,\rho^P]a\dg + b[\rho^P,\Pi_a\dg] + [\Pi_a, \rho^P]b\dg \big) \nn \\
&& + \sqrt{\etam} V(t) \bar{\mathcal{H}}[-\sqrt{\ko\etal} a + \sqrt{\kt}b]\rho^P + \sqrt{\etam} V(t) \bar{\mathcal{H}}[-\sqrt{\ko\etal} \Pi_a + \sqrt{\kt} \Pi_b]\rho^P \nn \\
&& +  \sum_{i=a,b} \Gamma_{\varphi}^i \mathcal{D}[\sigma_z^i]\rho +  \sum_{i=a,b} \Gamma_r^i \mathcal{D}[\sigma_-^i]\rho
\label{eq:sme_polaron}
\eqa
where $\bar{\mathcal{H}}[A]B \equiv AB+BA\dg$, $\Gamma_{1(2)} = \equiv \kappa_{\rm s,1(2)} + \kappa_{\rm w,1(2)}$ and the projectors are
\beq
\Pi_a \equiv (\Pi_{10}+\Pi_{11}) A^{(1)}(t) + (\Pi_{00}+\Pi_{01}) A^{(0)}(t), ~~~~~~ \Pi_b \equiv \Pi_{11} B^{(11)}(t) + \Pi_{10}B^{(10)}(t) + \Pi_{01}B^{(01)}(t) + \Pi_{00}B^{(00)}(t)
\eeq
and the Hamiltonian is
\beq
H^P =  -\frac{\omega_{q,1}}{2}\sigma_z^1 - \frac{\omega_{q,2}}{2} \sigma_z^2 + \omega_{r,1} a\dg a + \omega_{r,2} b\dg b + \chi_1 a\dg a \sigma_z^1 + \chi_2 b\dg b \sigma_z^2
\eeq
\\
\\
Notice that in this frame there is no drive of the cavity modes because we are dynamically shifting the cavity states back to the vacuum. Therefore, if the cavity starts off in the vacuum state it always remains in the vacuum state in this frame. This makes simulation of the system much easier since we can drop all the terms above that contain a field operator (notice that all field operators in the dynamical equation are normally ordered so that they annihilate the vacuum). This leaves us with an equation of motion just for the qubits that we can normalize and simulate efficiently. However, at the final time we must transform back into the lab frame from the polaron frame in order to interpret the results consistently. This transformation can be calculated easily by noting that  $\rho(t) = U\dg(t) \rho^P(t) U(t)$. Consider a general state in the polaron frame:
\beq
\varrho^P(t) = \sum_{ijkl} r_{ijkl}(t) \ket{ij}_q\bra{kl} \otimes \ket{00}_c\bra{00}
\eeq
where the first two components are qubit states (indicated by the subscript $q$) and the second two components are the cavity states (indicated by the subscript $c$). As before we can assume that in the polaron frame the cavity states remain the vacuum. We want to transform back into the lab frame and then trace over the cavity states. That is,
\beq
\rho(t) = \tr_{c}\left( U(t) \rho^P(t) U\dg (t) \right)
\eeq
where the trace is over cavities 1 and 2. Since in the current work the cavities at at their steady states at the "final time" (when the state characterization is done), we will specialize to the case where the operator $U(t) \rightarrow U_{ss}$ is time-independent because the states $A^{(i)}$ and $B^{(ij)}$ are in their steady state (which we indicate by dropping the time index). In this case, \bqa
\rho(t) &=& \tr_{c}\left( U_{ss} \rho^P(t) U_{ss}\dg \right) =  \sum_{ijkl} r_{ijkl}(t) \tr \left( U_{ss}\ket{ij}_q\bra{kl} \otimes \ket{00}_c\bra{00} U\dg_{ss} \right) \nn \\
&=& r_{1111}(t) \ket{11}_q\bra{11} + r_{1010}(t) \ket{10}_q\bra{10} + r_{0101}(t) \ket{01}_q\bra{01}+ r_{0000}(t) \ket{00}_q\bra{00}  \nn \\
&+& \bigg\{ r_{1110}(t)  _c\bra{00}D\dg_1\left[A^{(1)}\right]D\dg_2\left[B^{(10)}\right]D_1\left[A^{(1)}\right]D_2\left[B^{(11)}\right]\ket{00}_c \ket{11}_q\bra{10} + \nn \\
&&~~ r_{1101}(t)  _c\bra{00}D\dg_1\left[A^{(0)}\right]D\dg_2\left[B^{(01)}\right]D_1\left[A^{(1)}\right]D_2\left[B^{(11)}\right]\ket{00}_c \ket{11}_q\bra{01} + \nn \\
&&~~ r_{1100}(t)  _c\bra{00}D\dg_1\left[A^{(0)}\right]D\dg_2\left[B^{(00)}\right]D_1\left[A^{(1)}\right]D_2\left[B^{(11)}\right]\ket{00}_c \ket{11}_q\bra{00} + \nn \\
&&~~ r_{1001}(t)  _c\bra{00}D\dg_1\left[A^{(0)}\right]D\dg_2\left[B^{(01)}\right]D_1\left[A^{(1)}\right]D_2\left[B^{(10)}\right]\ket{00}_c \ket{10}_q\bra{01} + \nn \\
&&~~ r_{1000}(t)  _c\bra{00}D\dg_1\left[A^{(0)}\right]D\dg_2\left[B^{(00)}\right]D_1\left[A^{(1)}\right]D_2\left[B^{(10)}\right]\ket{00}_c \ket{10}_q\bra{00} + \nn \\
&&~~ r_{0100}(t)  _c\bra{00}D\dg_1\left[A^{(0)}\right]D\dg_2\left[B^{(00)}\right]D_1\left[A^{(0)}\right]D_2\left[B^{(01)}\right]\ket{00}_c \ket{01}_q\bra{00} + h.c. \bigg\} 
\eqa
So we see that the diagonal elements are not effected by the transformation back to the lab frame, but that the off-diagonal elements are all scaled by additional factors. These factors can be easily worked out for steady state values of the cavity fields, for example,
\beq
\bra{00}D\dg_1\left[A^{(0)}\right]D\dg_2\left[B^{(01)}\right]D_1\left[A^{(1)}\right]D_2\left[B^{(11)}\right]\ket{00}  = \exp \left\{ i\im{A^{(0)*}A^{(1)}} + i\im{B^{(01)*}B^{(11)}} - \frac{|\delta_{10}|^2}{2} - \frac{|\Delta_{1101}|^2}{2} \right\}
\eeq
where $\delta_{ij} \equiv A^{(i)} - A^{(j)}$ and $\Delta_{ijkl} \equiv B^{(ij)}-B^{(kl)}$.
So we can propagate the system in the polaron frame (which is more efficient since the cavity states stay at the vacuum) and then at the end scale the off-diagonal elements to get the density matrix in the lab frame.

\section{Full Details of Experimental Setup}

\begin{figure}
\includegraphics[scale=0.6]{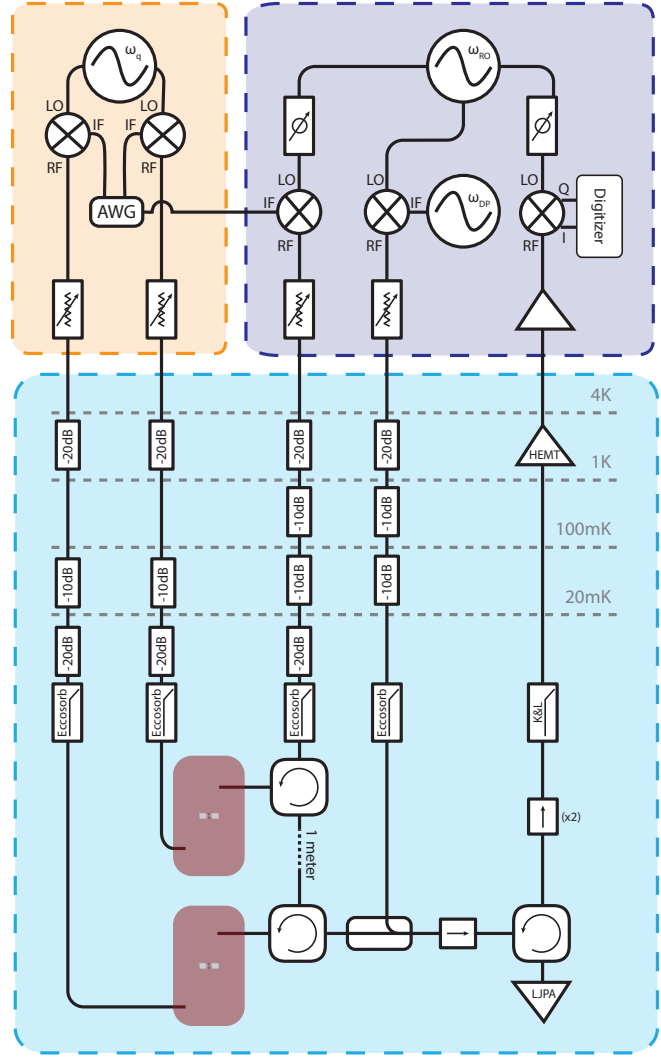}
\caption{Full experimental setup.}
\label{figure1}
\end{figure}

The joint measurement process described in this Letter requires the use of two GHz  microwave generators (to act as local oscillators for qubit and readout pulses); one MHz generator (for double-pumping the lumped-element Josephson parametric amplifier, LJPA); three DC current sources (for biasing the qubits and the amplifier); and an arbitrary waveform generator (AWG, for shaping qubit and readout pulses).  The full room- and base-temperature setup is shown in Figure  \ref{figure1}.  The qubits are housed at the base stage of a Vericold cryogen-free dilution refrigerator.  Input lines contain 50-60 dB of attenuation and additional homemade lossy Eccosorb low-pass filters at base stage to filter stray infrared radiation.  The qubits and cavities are housed in a blackened copper can, and the cavities are themselves indium-sealed to provide additional infrared shielding. Magnetic shielding is provided by wrapping the cavities individually with aluminum foil and by a $\mu$-metal outer shield that encompasses the copper can.

To implement qubit pulses, the qubits are first tuned to an operating frequency of $\omega_{q,1}/2\pi = 4.31143$ GHz and $\omega_{q,2}/2\pi= 4.46143$ GHz.  Qubit pulses are implemented using single-sideband modulators (SSBs) with the output of a first generator operating at the midpoint of the two qubit frequencies $\omega_{q}/2\pi = 4.38643 $ GHz serving as the local oscillator (LO).  The AWG provides intermediate frequency (IF) pulses at 75 MHz to a lower-sideband SSB (qubit 1) and an upper-sideband SSB (qubit 2); these pulses are routed to base and perform single qubit gates via the weakly-coupled ports of the respective cavities. 
\\
\\
A second generator operating at the measurement frequency $\omega_m/2\pi = 7.19326$ GHz is split three ways.  The joint measurement readout pulses are implemented via a mixer using $\omega_m$ as the LO and DC pulses from the AWG as the IF.  The ouput of the mixer passes through a variable phase shift and attenuation and into the dilution refrigerator.  At base, the readout passes through a circulator to measure cavity 1 in reflection; is routed back through the circulator and through 1.3 meters of copper cable; measures cavity 2 in reflection via a second circulator; and is routed via an additional isolator to an LJPA for phase-sensitive amplification.  We double-pump the LJPA symmetrically at $\omega_m \pm \omega_{dp}$ to reduce signal leakage at $\omega_m$.  The double pump for the LJPA is generated via an IQ mixer with a third generator operating at $\omega_{dp}/2\pi=369$ MHz on the I port and a $\omega_m$ as the LO.  The amplified readout passes through two isolators and a low-pass filter en route to a 4K HEMT; at room temperature, the signal is further amplified before demodulation (using the third branch of $\omega_m$ as the LO) and digitization for processing.

\section{Choosing an Operating Frequency}

As noted in the first section of this Supplement, the output of a double-reflection measurement at measurement frequency $\omega_m$ for two 3D transmons with bare cavity frequencies $\omega_{r,1}$ and $\omega_{r,2}$, measurement port bandwidths of $\kappa_{1,2}$, and dispersive shfits $\chi_{1,2}$ is given by the product of two complex reflection coefficients.  At steady state, we find:
\begin{equation}
B_{out} = \sqrt{\eta_{loss}}\frac{ \frac{\kappa_{s,1}}{2} - i\left(\omega_{r,1} - \omega_{m} \pm \chi_1 \right)} {\frac{\kappa_{s,1}}{2}+ i \left(\omega_{r,1}-\omega_{m} \pm \chi_1\right) } \times
 \frac{ \frac{\kappa_{s,2}}{2} - i\left(\omega_{r,2} -\omega_{m} \pm \chi_2 \right)} {\frac{\kappa_{s,2}}{2} + i \left(\omega_{r,2}-\omega_{m} \pm \chi_2\right) }A_{d}
\end{equation}
For qubits that are red-detuned from the cavities, $+(-)$ corresponds to a qubit in $|0\rangle(|1\rangle)$.  We have assumed that $Q_{int} \gg Q_{ext}$, such that we can neglect internal losses.  This leads to four distinct resonance curves for the states $|00\rangle$, $|01\rangle$, $|10\rangle$, and $|11\rangle$ (Figure~S 2).
Solving for the frequency at which $B_{out}^{(10)}= B_{out}^{(01)}$ results in a quadratic equation in $\omega_m$ that has real solutions if the following inequality is satisfied:
\begin{equation}
\left (\omega_{r,1} - \omega_{r,2} \right)^2 \geq \left( \frac{1}{4} \frac{\kappa_1 \kappa_2}{\chi_1 \chi_2} + 1 \right) \left[ \left( \chi_1 - \chi_2 \right)^2 - \frac{\chi_1\chi_2}{\kappa_1\kappa_2} \left( \kappa_1 - \kappa_2 \right)^2 \right].
\end{equation}
Careful manufacture of qubits and cavities enables us to match $\kappa_1$ and $\kappa_2$ within 2-3 MHz, and $\chi_1$ and $\chi_2$ within several hundred kHz.
As a result, this condition is fairly straightforward to meet by adjusting the cavity frequencies such that $|\omega_{r,1}-\omega_{r,2}| \sim \kappa$.
It is possible to theoretically calculate the correct $\omega_m$; in practice, we sequentially prepare the computational states $|00\rangle$, $|01\rangle$, $|10\rangle$, and $|11\rangle$, and adjust $\omega_m$ until the single-shot Gaussian measurement histograms for the $|01\rangle$ and $|10\rangle$ states completely overlap, as shown in Figure 2 of the main paper. 
\begin{figure}
\begin{center}
\includegraphics[scale=0.6]{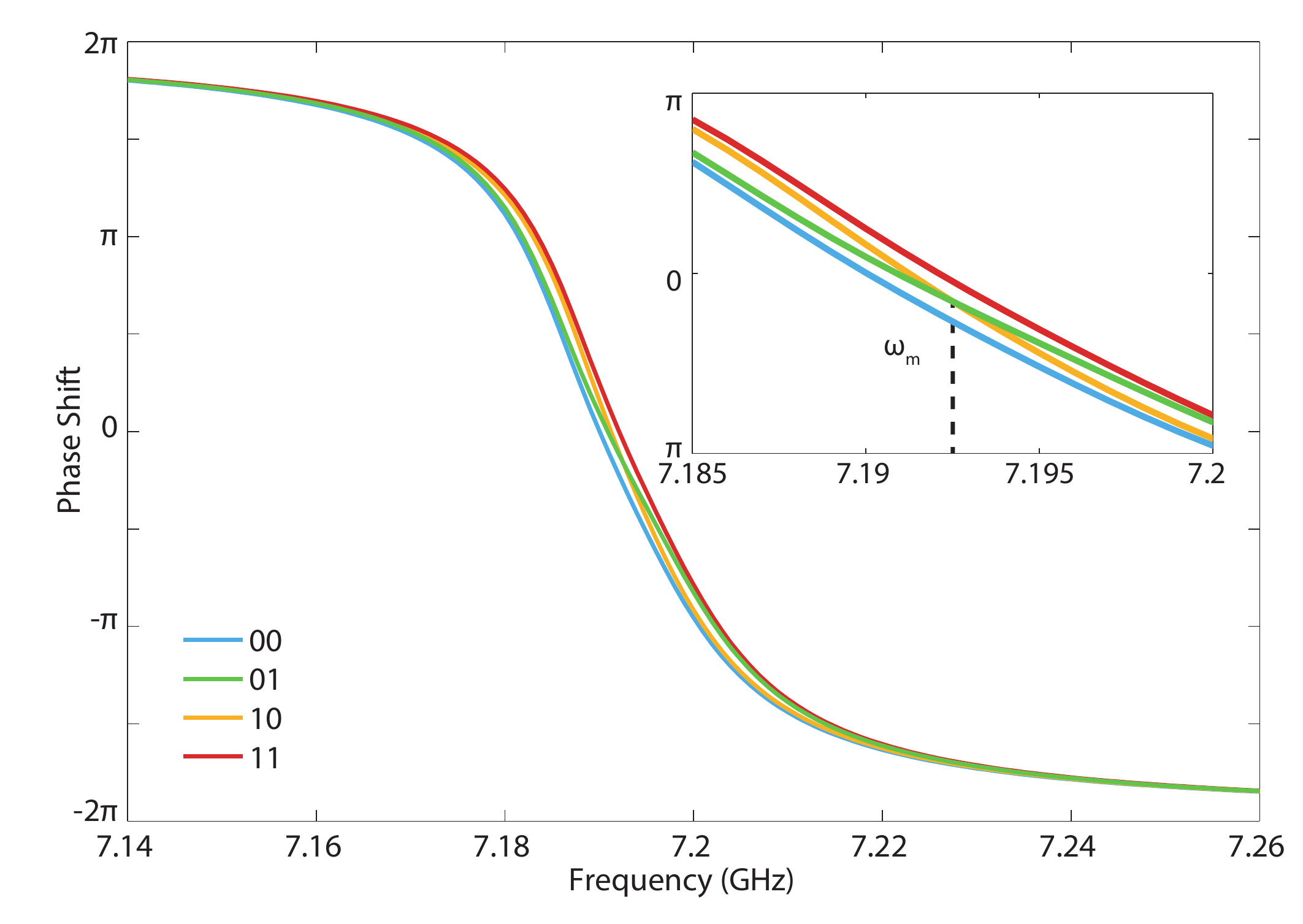}
\caption{Double-reflection phase shift calculated for the four prepared states $|00\rangle$, $|01\rangle$, $|10\rangle$, and $|11\rangle$.  The reflection curves pass through a $4\pi$, indicating reflection from two sequential cavities.  The inset shows the crossing between the reflected phases for $|01\rangle$ and $|10\rangle$ at $\omega_m = 7.19326$ GHz.}
\end{center}
\label{figureS2}
\end{figure} 

\section{System Calibration}

In order to effectively model our two-qubit system, we need to fully characterize it.  The necessary calibrations include: qubit lifetime and dephasing time $T_1$ and $T_2$; bare cavity frequencies $\omega_r$ and linewidths $\kappa$; dispersive shifts $\chi$; amplification efficiency $\eta_{meas}$ and inter-cavity transmission efficiency $\eta_{loss}$; measurement photon number $\overline{n_1}$; and gain of the amplification chain $G_{chain}$.  The calibrated values of these parameters are given in Table S1.  The calibration methods are described below.
\subsection{Cavity Frequencies ($\omega_r$) and Linewidths ($\kappa$)}
The cavity frequencies $\omega_{r,i}+\chi_i$ (that is, the cavity frequency for qubit $i$ in $|0\rangle$) and $\kappa_i$ are determined via $S_{21}$ measurements at base temperature.  

\subsection{Qubit Lifetimes and Coherences}
We calibrate $T_1$ and $T_2^*$ using standard time-resolved measurements.  

\subsection{Photon Number ($\bar{n}_1$), Dispersive Shifts ($\chi$) and Inter-Cavity Transmission Efficiency ($\eta_{loss}$)}
To calibrate $\chi$, $\eta_{loss}$ and $\bar{n}_1$, we use a technique similar to~\cite{Vijay}. We fix a measurement frequency $\omega_m$ and perform Ramsey measurements with an additional constant input power $P_m$ (calibrated at room-temperature with a spectrum analyzer), which corresponds to a coherent state in the cavities given by
\begin{equation}\alpha^{\pm} = \frac{\sqrt{\lambda P_m \kappa}}{\kappa/2 + i\left(\omega_{r} - \omega_m \pm \chi \right)}, \end{equation}
where $\lambda$ represents an unknown (but constant at fixed frequency) attenuation from room-temperature to the cavity.  All variables but $\chi$ and $\lambda$ are independently calibrated.  The intracavity coherent state $\alpha$ creates a measurement-induced dephasing rate given by $\Gamma_m = \frac{\kappa}{2}|\alpha^+-\alpha^-|^2$ and an ac-Stark shift of $\Delta_\omega = -2\chi Re\left[ \alpha^- \left(\alpha^{+}\right)^* \right]$\cite{Gambetta-08}.  The frequency of the Ramsey fringes gives $\Delta_\omega$, and the exponential decay envelope gives $\Gamma_{tot} = \Gamma_m+1/T^*_2$.  Taking the ratio $\Delta_\omega / \Gamma_{tot} \sim \Delta_\omega / \Gamma_m$ removes dependency on $\lambda P_m$; we fit this ratio to a constant from which we extract $\chi$.  We then use this value of $\chi$ in linear fits of $\Delta_\omega(P_m)$ and $\Gamma_m(P_m)$; this provides two independent fits for $\lambda$. In Figure 3, we show linear fits for $\Delta_\omega(P_m)$ and $\Gamma(P_m)$ for both qubits, in both instances using readout via the double-reflection measurement such that $\lambda_2 = \lambda_1 \eta_{loss}$.  This provides our calibration of $\eta_{loss}$. The calibration of $\lambda$ also provides a sensitive photon-number calibration as a function of $P_m$: once $\chi$ and $\lambda$ have been determined, $\overline{n}^{\pm} = |\alpha^{\pm}|^2$ is fully determined.

\subsection{Amplification Efficiency ($\eta_{meas}$) and Gain of the Amplification Chain ($G_{chain}$)}

$G_{chain}$ links the digitized measurement voltage $V_m(t)$ definied in the main paper, and $B_{out}$ as defined in the paragraph "Simplified Theory". $G_{chain}$ is thus the slope of the line $V_m^{(11)} -V_m^{(00)}$ vs $S_{11}-S_{00}$, where $V_m^{ij}$ is the center of the histogram actually measured with the digitizer (Figure S 4), and $S_{ij}$ is defined in Equation \ref{S_ij}. Once $G_{chain}$ is determined, we fit the histograms corresponding to the prepared state 00, 01, 10 and 11 (similar to the ones shown in Figure 2 of the main paper) to Gaussian distributions for every measurement time $t_m$. The amplification efficiency $\eta_{meas}$ is linked to the standard deviation of these gaussians via $1/\sigma=2\sqrt{\eta_{meas}t_m}$. We extract $\eta_{meas}$ by fitting $\sigma$ vs $1/\sqrt{t_m}$ to a line for every prepared state (See Figure S 4). We define $\eta_{meas}$ as the mean of these extracted values.

\begin{figure}
\begin{center}
\includegraphics[scale=0.65]{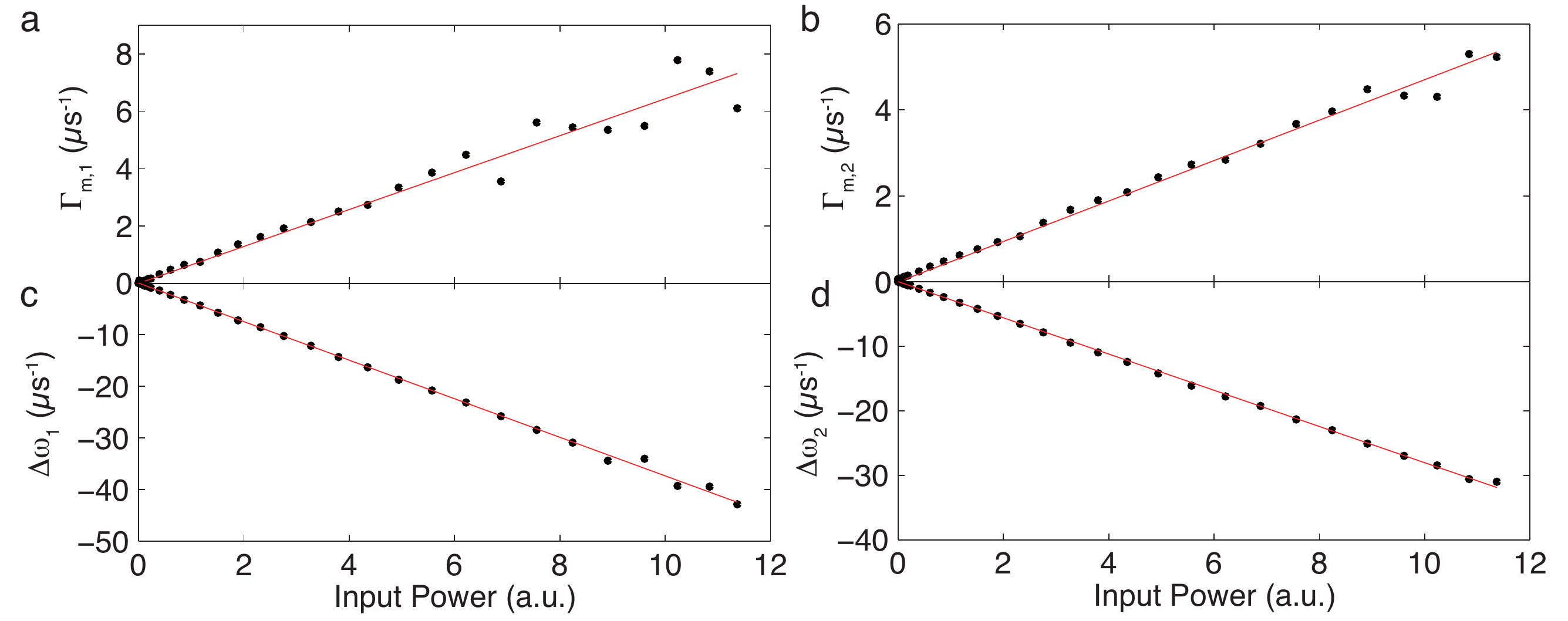}
\caption{Measurement-induced dephasing (a-b) and ac-Stark shift (c-d) for qubit 1 (left) and qubit 2 (right) as a function of measurement power $P_m$ in arbitrary units.}
\end{center}
\end{figure} 

\begin{figure}
\begin{center}
\includegraphics[scale=0.65]{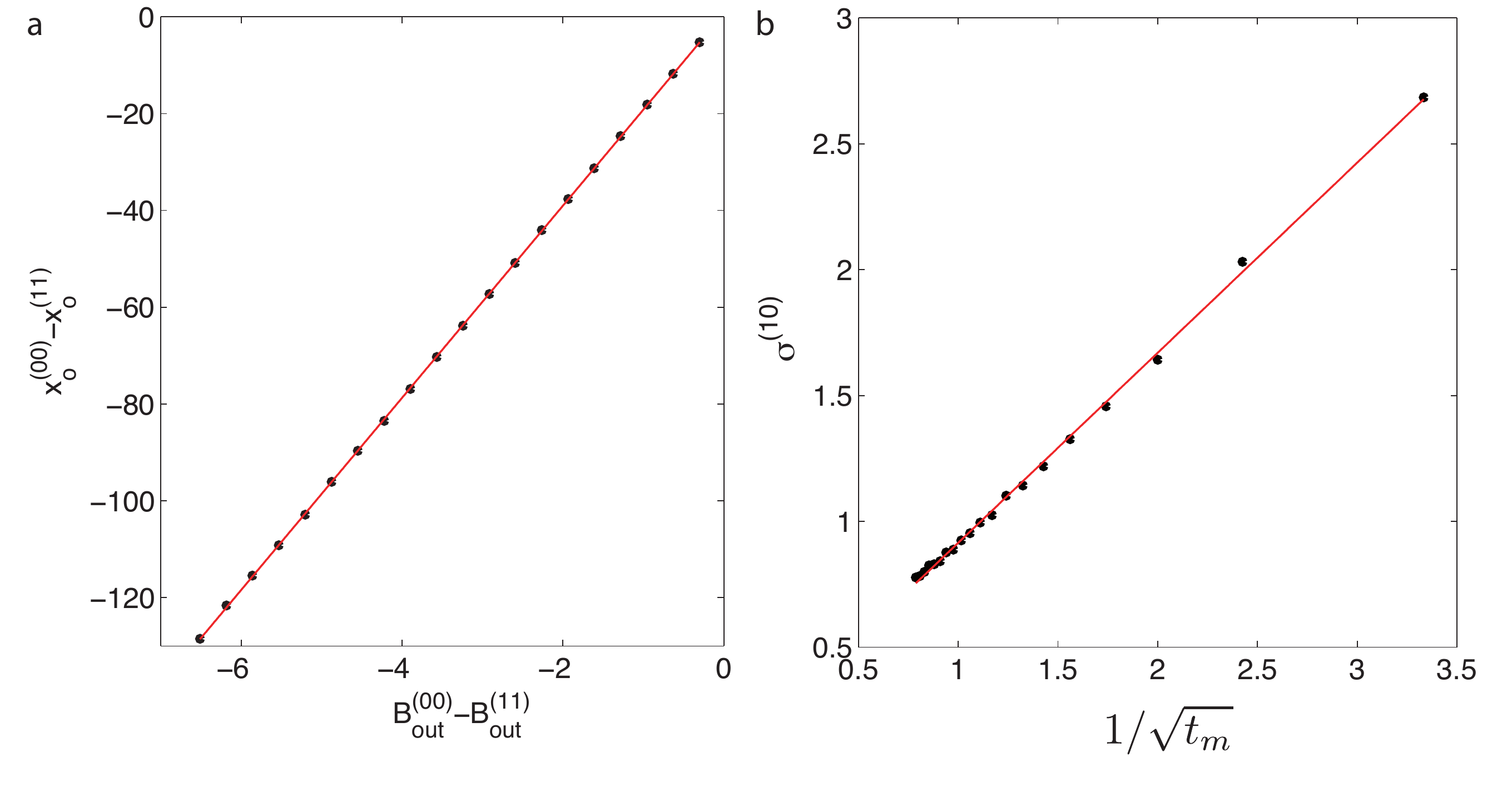}
\caption{a. Center of the measured histograms as a function of prediction given by the "simplified theory". We plot the difference between 00 and 11 to get rid of possible offsets. b. Evolution of the standard deviation of the histogram corresponding to prepared state 10 versus measurement time $t_m$.}
\end{center}
\end{figure} 

\begin{center}
\begin{tabular} {|c| c| c|}
\hline
\multicolumn{3}{|c|}{\textbf{Table S1: System Parameters}} \\ \hline
			& Qubit 1 		& Qubit 2 		\\ \hline
$\omega_q/2\pi$	& 4.31143 GHz	& 4.46143 GHz 	\\ \hline
$\omega_{r}/2\pi$	&7.1864 GHz		&7.1984 GHz		\\ \hline
$\kappa/2\pi$		&18.5 MHz		&21 MHz		\\ \hline
$\chi/2\pi$		&$1.275\pm0.025$ MHz		&$1.085\pm0.035$ MHz 		\\ \hline
$T_1$			& $27\pm 5$ $\mu$s 		&$20\pm 3$ $\mu$s 		\\ \hline
$T_2^*$		&$16\pm3$ $\mu$s 		&$12\pm2$ $\mu$s 		\\ \hline
$\eta_{loss}$		&\multicolumn{2}{|c|}{$0.81\pm0.05$} 			\\ \hline
$\eta_{meas}$	&\multicolumn{2}{|c|}{$0.4\pm0.10$}			\\ \hline
$G_{chain}$ &\multicolumn{2}{|c|}{$19.8\pm1.6$} \\ \hline
%\caption{Table S1: Calibrated system parameters}

\end{tabular}
\end{center}

\section{Tomography}

\begin{figure}
\begin{center}
\includegraphics[scale=0.6]{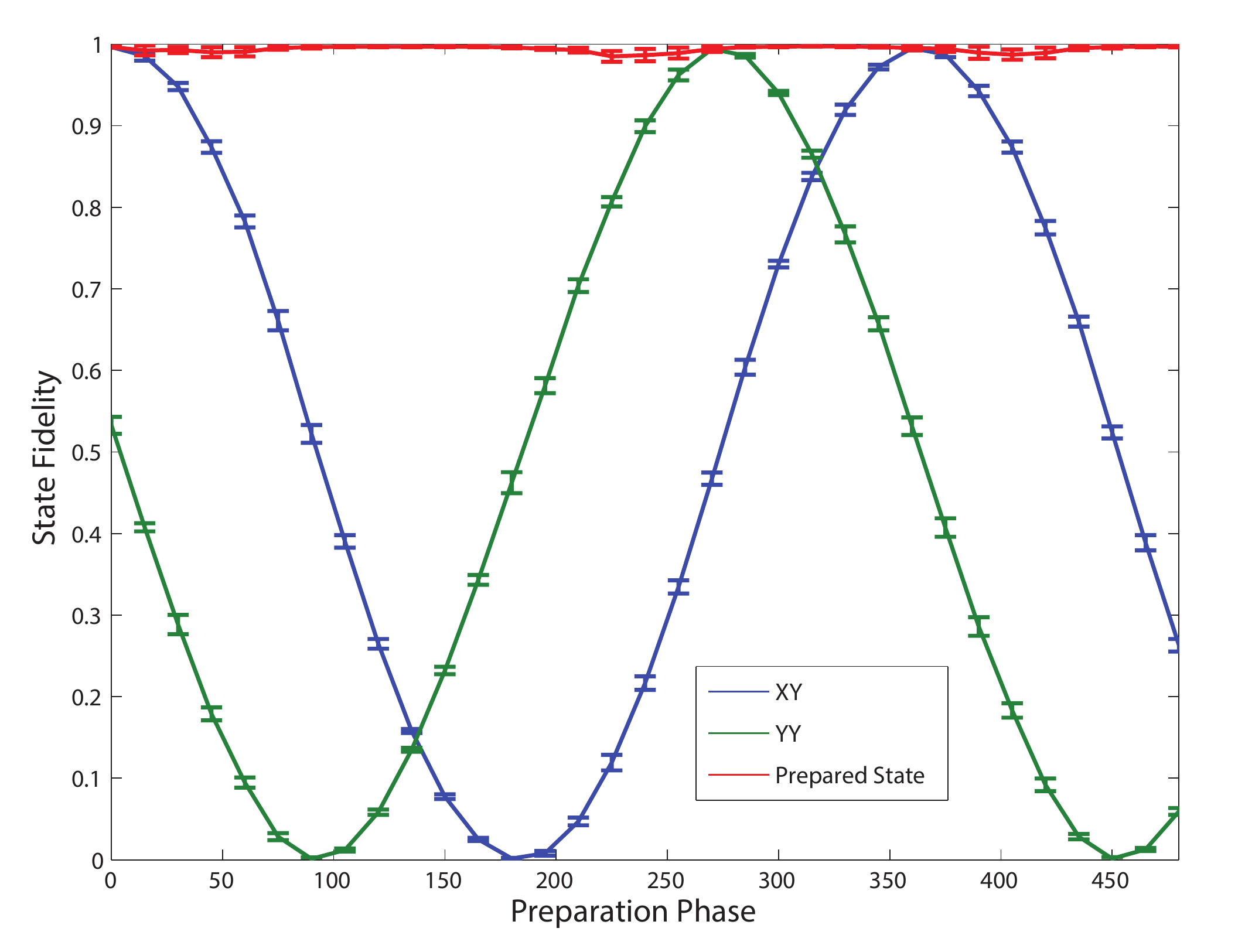}
\caption{Fidelity to the states $\frac{|0\rangle + |1\rangle}{\sqrt{2}} \bigotimes \frac{|0\rangle +|1\rangle}{\sqrt{2}}$ (XX), $\frac{|0\rangle +|1\rangle}{\sqrt{2}} \bigotimes \frac{|0\rangle + i|1\rangle}{\sqrt{2}}$ (XY) and $\frac{|0\rangle +|1\rangle}{\sqrt{2}} \bigotimes \frac{|0\rangle +e^{i\phi}|1\rangle}{\sqrt{2}}$ (the Target State) as a function of Qubit 2 preparation phase $\phi$.  Fidelities to XX and XY oscillate 90 degrees out of phase with one another, as expected; the fidelity to the prepared state is an average of 98.8\% across all preparation angles.}
\end{center}
\end{figure} 

To tomographically reconstruct the density matrix, we need at least fifteen linearly independent measurements in order to fully determine the 15 degrees of freedom of the two-qubit density matrix.  Our tomography procedure utilizes an overspecified set of 30 qubit rotations (15 positive and negative rotations) in order to reduce systematic bias from qubit rotations and power drifts.  The rotations are identical to those in Chow et al. \cite{Chow}.  We take advantage of the single-shot nature of our readout to extract the probabilities $p_{|00\rangle}$ and $p_{|11\rangle}$ for each of the 30 rotations, thus doubling the information we gather about the joint qubit state for each measurement.  These probabilities represent measurements of the form $\beta_{II}\sigma_{II} \pm \beta_{IZ} \sigma_{Im} \pm \beta_{ZI} \sigma_{nI} + \beta_{ZZ} \sigma_{nm}$, where $+(-)$ corresponds to $p_{|00\rangle}\left(p_{|11\rangle}\right)$, and $m,n \in \{ X,Y,Z\}$. The $\beta$-coefficients are calibrated using a double-Rabi measurement as described in Chow \textit{et al}\cite{Chow}. Our measurement set results in an overspecified measurement set: 60 measurements for 15 degrees of freedom.  We convert this data into a density matrix using a least-squares maximum likelihood estimation method to enforce trace normalization and Hermiticity of the reconstructed density matrix.  We verify the accuracy of the tomography by preparing the state $\frac{|0\rangle + |1\rangle}{\sqrt{2}} \bigotimes \frac{|0\rangle + e^{i\phi}|1\rangle}{\sqrt{2}}$ and calcluating the fidelity of the resulting density matrix to the target state (Figure S5).  The average fidelity across the prepared states is 98.8\%, indicating highly effective state initialization and tomographic reconstruction.

\section{Reconstruction of single quantum trajectories}

Given a continuous measurement record $V_m(t)$, it is possible to reconstruct the quantum trajectory (time-dependent conditional density matrix) corresponding to it in several ways, as demonstrated in figure 4 of the main paper. In this section we present more details on the various methods for reconstruction.

\subsection{Reconstruction based on Bayesian update}

The Bayesian updating protocol draws on the simplified theory developed in Section 1 of this supplement.
For each measurement time $t_m$, it is possible to infer the density matrix of the two-qubit system by using a Bayesian update based on the measured integrated voltage $V_m(t_m)$ (See for example figure 2.a of the main paper). We first calculate $S_{ij}$ and $\sigma$ that we recall from Section 1: 

\be
  S_{ij}=\frac{1}{t}\int {\rm Re} [ B_{out}^{(ij)}(t') \, e^{-i\phi}] \,
   f_w(t') \, dt',
    \label{S_ij_2}\ee
where $f_w(t)$ is the weight function (in the experiment we used constant-weight integration with adjustable start/end time moments).  The amplifier noise is also accumulated during this time-integration, so that for the two-qubit state $|ij\rangle$ the random measurement result is characterized by the Gaussian distribution with the mean value of $S_{ij}$ and the standard deviation
    \be
    \sigma = \frac{1}{2\sqrt{\eta_{meas}}} \sqrt{\frac{1}{t}\int f_w^2(t)\, dt},
    \label{sigma_2}\ee

The selection probability for the initial state $|ij\rangle$ is then given by

    \be
  p_{sel}(i,j)=\int_{V_m-\delta V_m}^{V_m+\delta V_m} e^{-\frac{(V-S_{ij})^2}{2\sigma^2}} dV
    \label{p_sel}\ee

Density matrix elements can then be deduced using equations (\ref{rho_fin}) and (\ref{rho-01,10-num}).

\begin{figure}
\begin{center}
\includegraphics[scale=0.45]{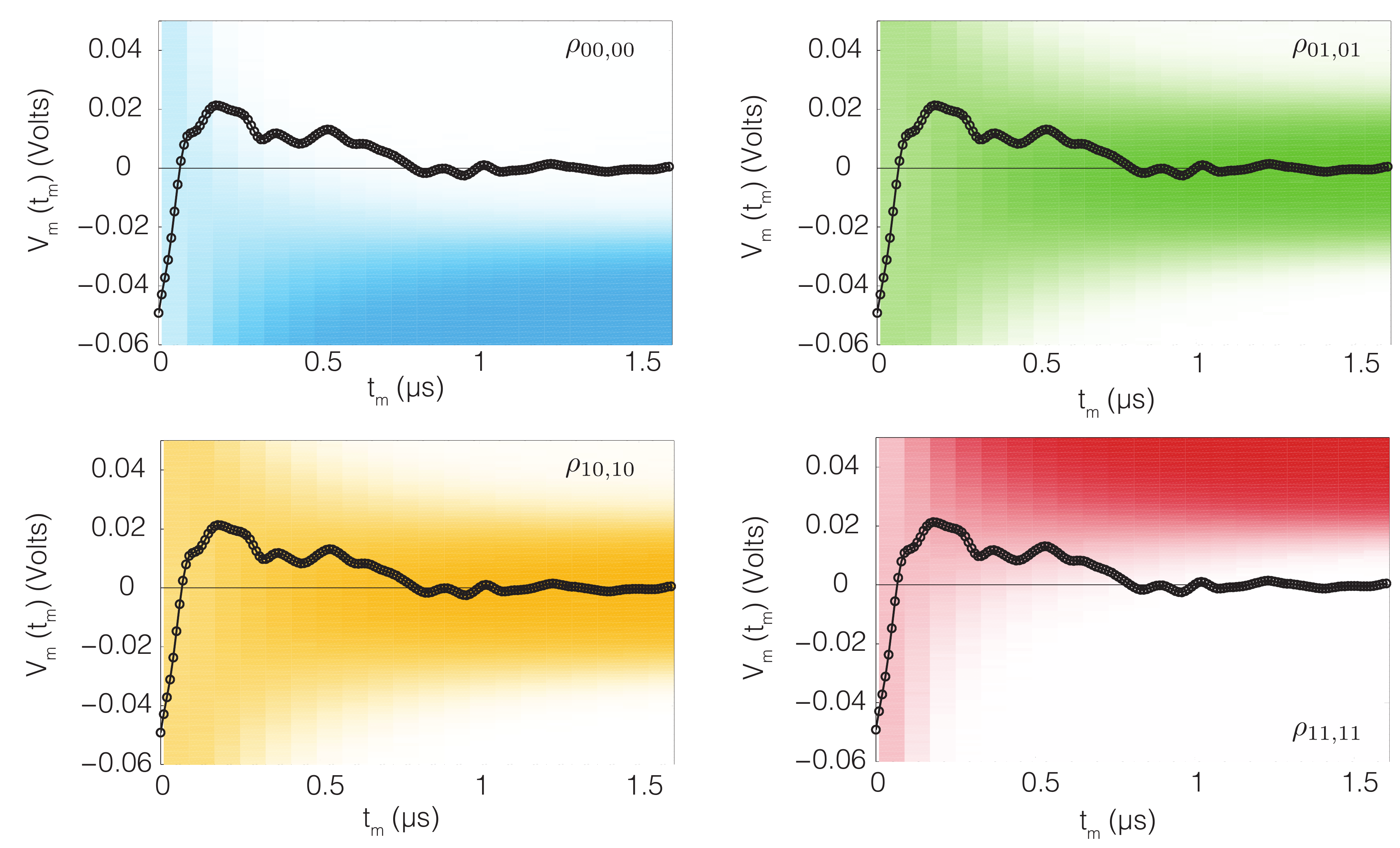}
\caption{Diagonal elements of the two-qubit density matrix conditioned on $V_m(t_m)$ and $t_m$, obtained for $\bar{n}=1.2$. The color code ranges from 0 (white) to 1 (maximum intensity color). That is, the intensity encodes the estimated value of the density matrix element for a given integrated voltage $V_m$ and measurement time $t_m$. The dotted line is an example of the temporal evolution of the measurement signal $V_m$ similar to the one on the figure 2.a of the main paper.}
\label{condtomo_diag}
\end{center}
\end{figure} 

\begin{figure}
\begin{center}
\includegraphics[scale=0.45]{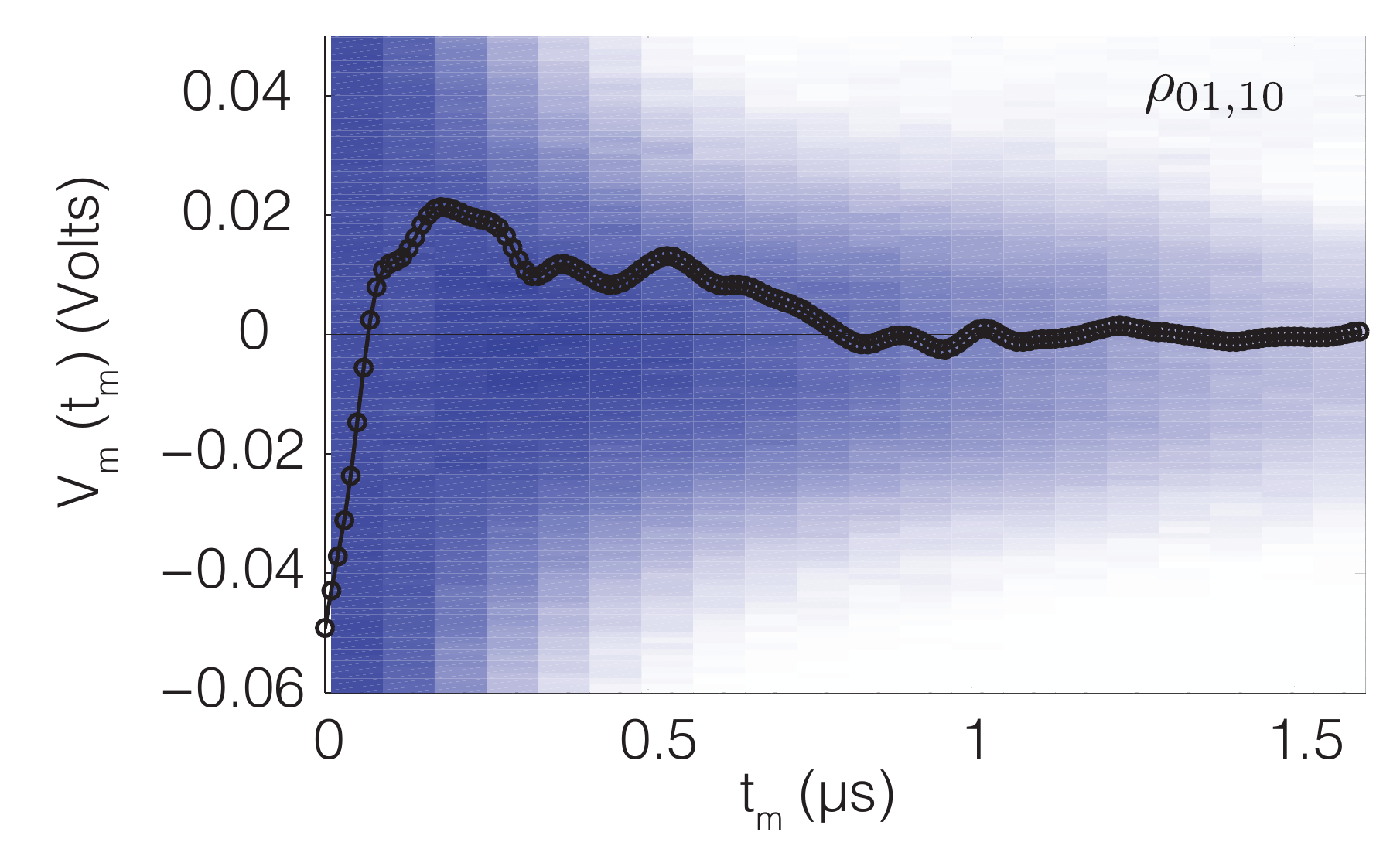}
\caption{Absolute value of the off-diagonal element $\rho_{01,10}$ of the two-qubit density matrix conditioned on $V_m(t_m)$ and $t_m$, obtained for $\bar{n}=1.2$. The color code ranges from 0 (white) to 1 (maximum intensity color). That is, the intensity encodes the estimated value of the density matrix element for a given integrated voltage $V_m$ and measurement time $t_m$. The dotted line is an example of the temporal evolution of the measurement signal $V_m$ similar to the one on the figure 2.a of the main paper.}
\label{condtomo_offdiag}
\end{center}
\end{figure} 

\subsection{Experimental reconstruction}

This reconstruction is based on the ability to map $V_m$ to the actual density matrix. This mapping $V_m \mapsto \rho \left(  V_m \right)$ is obtained by performing conditional tomography for measurement outcomes lying within the window $[V_m-\delta V_m,V_m+\delta V_m]$ for each measurement time $t_m$ (See figures S\ref{condtomo_diag} and S\ref{condtomo_offdiag}). To obtain single quantum trajectories, we then just have to superimpose a single realisation of the measurement signal $V_m$ to this experimentally obtained map.

\subsection{Reconstruction based on stochastic master equation}

The conditional density matrix can also be recovered from a measurement trace $V_m(t)$ using the stochastic master equation in \erf{eq:sme_polaron}. One procedure for this reconstruction is very similar to the experimental reconstruction detailed above. Explicitly, \erf{eq:sme_polaron} is simulated for $100000$ instantiations of the Wiener noise, a time step of $1$ns, and a simulation time of $t_m$.  The homodyne voltage is obtained from the time average of the instantaneous voltage 
\beq
V(t) = \sqrt{\etam}\langle -\sqrt{\ko\etal} \Pi_a + \sqrt{\kt} \Pi_b] \rangle + \xi(t)
\eeq
and density matrices conditioned on realizations of the voltage within $[V_m-\delta V_m,V_m+\delta V_m]$ are summed and averaged to give estimates of the conditional density matrix elements.  

In principle, it is possible to directly drive the stochastic master equation with the measurement record (\ie  solve \erf{eq:sme_polaron} with the experimentally measured $V_m(t)$ substituted). However, for this to produce an accurate trajectory the temporal resolution of the measurement record has to be small ($<1$ns) since the stochastic master equation is derived in the infinitesimal time-step limit. This method was not possible in our case because the experimental apparatus used has a measurement resolution of $\sim 10$ns.

\end{document}